\documentclass[12pt]{iopart}

\usepackage{iopams}
\usepackage{graphicx}

\newcommand{\bmath}[1]{\mbox{{\boldmath{{$#1$}}}}}

\newtheorem{proposition}{Proposition}
\newtheorem{theorem}[proposition]{Theorem}
\newtheorem{lemma}[proposition]{Lemma}
\newtheorem{corollary}[proposition]{Corollary}

\begin{document}

\title[${\mathfrak{su}}(N)$ EYM solitons and black holes
with $\Lambda <0$]{On the existence of soliton and hairy black hole
solutions of ${\mathfrak {su}}(N)$ Einstein-Yang-Mills theory
with a negative cosmological constant}

\author{J. E. Baxter and Elizabeth Winstanley}

\address{School of Mathematics and Statistics,
The University of Sheffield,
Hicks Building,
Hounsfield Road,
Sheffield.
S3 7RH
United Kingdom}

\eads{E.Winstanley@sheffield.ac.uk}

\begin{abstract}
We study the existence of soliton and black hole solutions of four-dimensional
${\mathfrak {su}}(N)$ Einstein-Yang-Mills theory with a negative cosmological constant.
We prove the existence of non-trivial solutions for any integer $N$,
with $N-1$ gauge field degrees of freedom.
In particular, we prove the existence of solutions in which all the gauge field functions have no zeros.
For fixed values of the parameters (at the origin or event horizon, as applicable)
defining the soliton or black hole solutions, if the magnitude of the cosmological constant
is sufficiently large, then the gauge field functions all have no zeros.
These latter solutions are of special interest because at least some of them will be linearly stable.
\end{abstract}



\maketitle

\section{Introduction}
\label{sec:intro}

Since the discovery of non-trivial, spherically symmetric, soliton \cite{Bartnik1} and black hole \cite{Bizon}
solutions of the ${\mathfrak {su}}(2)$ Einstein-Yang-Mills (EYM) equations
in four-dimensional asymptotically flat space-time, there has been
an explosion of interest in the properties of solutions of EYM theory and its variants
(see, for example, \cite{Volkov1} for a review).
The original soliton solutions were a surprise discovery because there are no gravitational solitons
\cite{Heusler1}
nor solitons in pure Yang-Mills theory in flat space-time \cite{Deser1}.
The black hole solutions are counter-examples to the ``no-hair'' conjecture \cite{Wheeler1}, in that
they have no global charge, and are indistinguishable from a standard Schwarzschild black hole at infinity.
However, while the ``letter'' of the no-hair conjecture is violated, its ``spirit'' remains intact because
these black hole solutions, like the solitons, are linearly unstable \cite{Straumann1}.

Introducing a negative cosmological constant changes the picture completely (see \cite{Winstanley3}
for a recent review).
In asymptotically flat space, soliton and black hole solutions exist at discrete points in the parameter
space, but in four-dimensional asymptotically anti-de Sitter (adS) space,
${\mathfrak {su}}(2)$ EYM solutions are found in continuous open
sets of the parameter space \cite{Winstanley1,Bjoraker,Breitenlohner1}.
Of particular interest are those solitons and black holes for which the single gauge field function $\omega $
has no zeros.
In asymptotically flat space, all solutions are such that $\omega $ must have at least one zero,
and the number of unstable perturbation modes of the solutions is $2k$, where $k$ is the number of zeros
of $\omega $ \cite{Lavrelashvili1}.
In asymptotically adS space, at least some (but, interestingly, not all \cite{Breitenlohner1})
solutions where $\omega (r)$ has no zeros are linearly stable under spherically symmetric
\cite{Winstanley1,Bjoraker,Breitenlohner1} and non-spherically symmetric \cite{Sarbach}
perturbations, provided $\left| \Lambda \right| $ is sufficiently large.
Stable black holes in this model violate the  ``no-hair'' conjecture in that, at infinity, they are indistinguishable
from magnetically-charged Reissner-Nordstr\"om-adS black holes.
However, only one extra parameter (which can be taken to be the value of the gauge field function at
the event horizon) is required to completely determine the geometry exterior to the event horizon.
While this is not a global charge measurable at infinity, one might still argue that the ``spirit'' of the ``no-hair''
conjecture is still valid as only a small number of parameters are required to fully describe stable
black holes.

Recently soliton and black hole solutions of four-dimensional
${\mathfrak {su}}(N)$ EYM theory with a negative cosmological
constant have been found \cite{Winstanley3,Baxter1,Baxter2}.
For purely magnetic solutions, the gauge field is described by $N-1$ gauge field functions $\omega _{j}$.
As in the ${\mathfrak {su}}(2)$ case, there are solutions for which all the gauge field functions
have no zeros.
For $\left| \Lambda \right| $ sufficiently large, at least some of these are stable, for any value of $N$
\cite{Baxter1,Baxter3}.
Stable black hole solutions in ${\mathfrak {su}}(N)$ EYM in adS therefore require an additional $N-1$
parameters to completely describe the configuration, even though the metric is identical to magnetically
charged Reissner-Nordstr\"om-adS at infinity.
Therefore an unbounded number of parameters is required to determine the structure of stable hairy black holes
in adS.

Our purpose in the present paper is to prove analytically the existence of solitons and hairy black holes
in four-dimensional ${\mathfrak {su}}(N)$ EYM in adS, confirming the numerical work presented in \cite{Baxter2}.
Existence theorems for the original soliton and black hole
solutions of ${\mathfrak {su}}(2)$ EYM in asymptotically flat space were challenging to prove
\cite{Breitenlohner2,Smoller}, but it has been shown analytically in this case that there are an infinite number
of soliton and black hole solutions of the field equations, parameterized by the radius of the event horizon
(if there is one) and $k$, the number of zeros of the gauge field function $\omega $.
It has also been proven that $k>0$.
For ${\mathfrak {su}}(N)$ EYM in asymptotically flat space, rather less analytic work exists in the literature
\cite{Ruan,Winstanley2}.
Arguments for the existence of solutions for arbitrary $N$ are presented in \cite{Winstanley2}, but the only
complete analytic work is for ${\mathfrak {su}}(3)$ \cite{Ruan}.
For ${\mathfrak {su}}(2)$ EYM in adS, proving the existence of solutions \cite{Winstanley1} is considerably easier
than in the asymptotically flat case, mostly due to the boundary conditions at infinity.
It is therefore reasonable to imagine that the corresponding proof for ${\mathfrak {su}}(N)$ EYM in adS will also
be significantly easier than in the asymptotically flat case.

The outline of this paper is as follows.
We start, in Section \ref{sec:EYMtheory}, by outlining the field equations, gauge field ansatz and boundary conditions
for soliton and hairy black hole solutions of ${\mathfrak {su}}(N)$ EYM theory with a negative
cosmological constant in four space-time dimensions \cite{Baxter2}.
Some trivial solutions of the field equations are discussed in Section \ref{sec:su2embedded}, the most important
of which (for our later analysis) are embedded ${\mathfrak {su}}(2)$ solutions.
We also review the salient features of the numerical solutions presented in \cite{Baxter2}.
The field equations, presented in Section \ref{sec:ansatz}, are singular at the origin, black hole event horizon
(if there is one) and at infinity.
The first part of the proof of the existence of solutions is therefore to prove local existence of solutions
satisfying the appropriate boundary conditions (which are given in Section \ref{sec:boundary}) at these singular
points, and this is the subject of Section \ref{sec:local}.
It turns out that this is the most lengthy and technically intricate part of the whole existence proof, particularly
for local solutions near the origin.
Our method in Section \ref{sec:local} follows a well-established approach
\cite{Breitenlohner2,Winstanley2,Oliynyk1,Kunzle2} which involves casting the field equations in an appropriate form,
so that a standard theorem (Theorem \ref{thm:BFM} \cite{Breitenlohner2}) can be applied.
It is deriving a suitable form for the equations near the origin which is the most complicated part of the analysis.
The standard theorem then gives us local existence of solutions in a neighbourhood of the singular point
(i.e. the origin, event horizon or infinity), and an
important consequence of this theorem is that the solutions are analytic in the parameters which determine them.
This property of analyticity is central to the arguments, presented in Section \ref{sec:existence}, which
prove the existence of genuinely ${\mathfrak {su}}(N)$ soliton and black hole solutions of the field equations.

Our existence theorems are presented in Sections \ref{sec:result} and \ref{sec:largeL}.
In Section \ref{sec:result}
we show existence of ${\mathfrak {su}}(N)$ EYM solutions in a neighbourhood of any embedded ${\mathfrak {su}}(2)$
solution, and, as a corollary of this, the existence of ${\mathfrak {su}}(N)$ solutions for which all
the gauge field functions have no zeros for any value of $\left| \Lambda \right| $.
The behaviour of the solution space for large $\left| \Lambda \right| $ is discussed in Section \ref{sec:largeL},
and, in particular, we show that for any fixed values of the ``initial'' parameters (defined in Section \ref{sec:largeL})
determining the solution of the field equations, for sufficiently large $\left| \Lambda \right| $, the
solution generated by these parameters will be such that all the gauge field functions have no zeros.
Finally, our conclusions are presented in Section \ref{sec:conc}.

\section{${\mathfrak {su}}(N)$ Einstein-Yang-Mills theory}
\label{sec:EYMtheory}

In this section we will describe in detail our field equations and boundary
conditions for both soliton and black hole solutions.
We will also describe some simple embedded solutions of the theory, as well as reviewing
the key features of the numerical solutions studied in \cite{Baxter2}.

\subsection{Ansatz and field equations}
\label{sec:ansatz}

We consider four-dimensional ${\mathfrak {su}}(N)$ EYM theory
with a negative cosmological constant, described by the following action, in suitable units:
\begin{equation}
S_{\mathrm {EYM}} = \frac {1}{2} \int d ^{4}x {\sqrt {-g}} \left[
R - 2\Lambda - {\mathrm {Tr}} \, F_{\mu \nu }F ^{\mu \nu }
\right] ,
\label{eq:action}
\end{equation}
where $R$ is the Ricci scalar of the geometry and $\Lambda $ the cosmological constant.
Throughout this paper, the metric has signature $\left( -, +, +, + \right) $ and
we use units in which $4\pi G = 1 = c$.
In this article we focus on a negative cosmological constant, $\Lambda <0$.
Varying the action (\ref{eq:action}) gives the field equations
\begin{eqnarray}
T_{\mu \nu } & = & R_{\mu \nu } - \frac {1}{2} R g_{\mu \nu } + \Lambda g_{\mu \nu };
\nonumber \\
0 & = & D_{\mu } F_{\nu }{}^{\mu } = \nabla _{\mu } F_{\nu }{}^{\mu }
+ \left[ A_{\mu }, F_{\nu }{}^{\mu } \right] ;
\label{eq:fieldeqns}
\end{eqnarray}
where the YM stress-energy tensor is
\begin{equation}
T_{\mu \nu } = \, {\mbox {Tr}} \, F_{\mu \lambda } F_{\nu }{}^{\lambda } - \frac {1}{4} g_{\mu \nu }
{\mbox {Tr}} \, F_{\lambda \sigma} F^{\lambda \sigma } .
\label{eq:Tmunu}
\end{equation}
In equations (\ref{eq:action}--\ref{eq:Tmunu}) we have employed the usual Einstein summation convention,
where a summation over repeated indices is understood.
However, from now on all summations will be given explicitly, and this will be particularly important
in Section \ref{sec:local}.

In this paper we are interested in static, spherically symmetric soliton and black hole
solutions of the field equations (\ref{eq:fieldeqns}), and we write the metric in standard
Schwarzschild-like co-ordinates as:
\begin{equation}
ds^{2} = - \mu S^{2} \, dt^{2} + \mu ^{-1} \, dr^{2} +
r^{2} \, d\theta ^{2} + r^{2} \sin ^{2} \theta \, d\phi ^{2} ,
\label{eq:metric}
\end{equation}
where the metric functions $\mu $ and $S$ depend on the radial co-ordinate $r$ only.
In the presence of a negative cosmological constant $\Lambda <0$, we write the metric function $\mu $ as
\begin{equation}
\mu (r) = 1 - \frac {2m(r)}{r} - \frac {\Lambda r^{2}}{3}.
\label{eq:mu}
\end{equation}
We emphasize that we are considering only spherically symmetric black holes and not topological black holes
which have been found in the ${\mathfrak {su}}(2)$ case \cite{Radu1}.

The most general, spherically symmetric, ansatz for the ${\mathfrak {su}}(N)$ gauge potential is \cite{Kunzle1}:
\begin{equation}
A  =
{\cal {A}} \, dt + {\cal {B}} \, dr +
\frac {1}{2} \left( C - C^{H} \right) \, d\theta
- \frac {i}{2} \left[ \left(
C + C^{H} \right) \sin \theta + D \cos \theta \right] \, d\phi ,
\label{eq:gaugepot}
\end{equation}
where ${\cal {A}}$, ${\cal {B}}$, $C$ and $D$ are all $\left( N \times N \right) $ matrices and
$C^{H}$ is the Hermitian conjugate of $C$.
The matrices ${\cal {A}}$ and ${\cal {B}}$ are purely imaginary, diagonal, traceless and depend only
on the radial co-ordinate $r$.
The matrix $C$ is upper-triangular, with non-zero entries only immediately above the diagonal:
\begin{equation}
C_{j,j+1}=\omega_j (r) e^{i\gamma _{j}(r)},
\end{equation}
for $j=1,\ldots,N-1$.
In addition, $D$ is a constant matrix:
\begin{equation}
D=\mbox{Diag}\left(N-1,N-3,\ldots,-N+3,-N+1\right) .
\label{eq:matrixD}
\end{equation}
Here we are interested only in purely magnetic solutions, so we set ${\cal {A}} \equiv 0$.
We may also take ${\cal {B}}\equiv 0$ by a choice of gauge \cite{Kunzle1}.
From now on we will assume that all the $\omega _{j}(r)$ are non-zero
(see, for example, \cite{Galtsov1}
for the possibilities in asymptotically flat space if this
assumption does not hold).
In this case one of the Yang-Mills equations becomes \cite{Kunzle1}
\begin{equation}
\gamma _{j} = 0 \qquad \forall j=1,\ldots , N-1.
\end{equation}
Our ansatz for the Yang-Mills potential therefore reduces to
\begin{equation}
A = \frac {1}{2} \left( C - C^{H} \right) \, d\theta - \frac {i}{2} \left[ \left(
C + C^{H} \right) \sin \theta + D \cos \theta \right] \, d\phi ,
\label{eq:gaugepotsimple}
\end{equation}
where the only non-zero entries of the matrix $C$ are
\begin{equation}
 C _{j,j+1} = \omega _{j}(r).
\end{equation}
The gauge field is therefore described by the $N-1$ functions $\omega _{j}(r)$.
We comment that our ansatz (\ref{eq:gaugepotsimple}) is by no means the only possible choice in
${\mathfrak {su}}(N)$ EYM.
Techniques for finding {\em {all}} spherically symmetric ${\mathfrak {su}}(N)$ gauge potentials
can be found in \cite{Bartnik2}, where all irreducible models are explicitly listed for $N\le 6$.

With the ansatz (\ref{eq:gaugepotsimple}), there are $N-1$ non-trivial Yang-Mills equations
for the $N-1$ gauge field functions $\omega _{j}$:
\begin{equation}
r^2\mu\omega''_{j}+\left(2m-2r^3 p_{\theta}-\frac{2\Lambda r^3}{3}\right)\omega'_{j}+W_j\omega_j=0
\label{eq:YMe}
\end{equation}
for $j=1,\ldots,N-1$, where a prime $'$ denotes $d /d r $,
\begin{eqnarray}
p_{\theta}&=&
\frac{1}{4r^4}\sum^N_{j=1}\left[\left(\omega^2_j-\omega^2_{j-1}-N-1+2j\right)^2\right],
\label{eq:ptheta}
\\
W_j&=&
1-\omega^2_j+\frac{1}{2}\left(\omega^2_{j-1}+\omega^2_{j+1}\right),
\label{eq:Wdef}
\end{eqnarray}
and $\omega_0=\omega_N=0$.
The Einstein equations take the form
\begin{equation}
m' =
\mu G+r^2p_{\theta},
\qquad
\frac{S'}{S}=\frac{2G}{r},
\label{eq:Ee}
\end{equation}
where
\begin{equation}
G=\sum^{N-1}_{j=1}\omega_j'^2.
\label{eq:Gdef}
\end{equation}
Altogether, then, we have $N+1$ ordinary differential equations for the $N+1$ unknown functions $m(r)$, $S(r)$
and $\omega _{j}(r)$.
The field equations (\ref{eq:YMe},\ref{eq:Ee}) are invariant under the transformation
\begin{equation}
\omega _{j} (r) \rightarrow -\omega _{j} (r)
\label{eq:omegaswap}
\end{equation}
for each $j$ independently, and also under the substitution:
\begin{equation}
j \rightarrow N - j.
\label{eq:Nswap}
\end{equation}


\subsection{Boundary conditions}
\label{sec:boundary}

We are interested in black hole and soliton solutions of the field equations
(\ref{eq:YMe},\ref{eq:Ee}).
However, the field equations are singular at the origin $r=0$ if we have a regular, soliton solution,
at an event horizon $r=r_{h}$ if there is one, and at infinity $r\rightarrow \infty $.
Since the cosmological constant $\Lambda <0$, there is no cosmological horizon.
We need to derive the boundary conditions at each of these singular points.
For $\Lambda =0$, local existence of solutions of the field equations near these singular points
has been rigorously proved \cite{Oliynyk1,Kunzle2}.
The extension of these results to $\Lambda <0$ will be the focus of Section \ref{sec:local}.

\subsubsection{Origin}
\label{sec:bcorigin}

The boundary conditions at the origin are the most complicated of the three singular points.
We postpone the detailed derivation of the form of the field variables near the origin to
Section \ref{sec:origin}, and instead here briefly state the results of this derivation for reference.

We assume that the field variables $m(r)$, $S(r)$ and $\omega _{j}(r)$ have regular Taylor series
expansions about $r=0$:
\begin{eqnarray}
m(r) & = & m_{0} + m_{1}r + m_{2}r^{2} + O(r^{3});
\nonumber \\
S(r) & = & S_{0} + S_{1}r +S_{2}r^{2} + O(r^{3});
\nonumber \\
\omega _{j}(r) & = & \omega _{j,0} + \omega _{j,1}r + \omega _{j,2} r^{2} + O(r^{3});
\label{eq:origin1}
\end{eqnarray}
where the $m_{i}$, $S_{i}$ and $\omega _{j,i}$ are constants.
The constant $S_{0}$ is non-zero in order for the metric to be regular at the origin,
but otherwise arbitrary since the field equations involve only derivatives of $S$.

Regularity of the metric and curvature at the origin immediately gives:
\begin{equation}
m_{0}=m_{1}=m_{2} =0, \qquad S_{1}=0, \qquad \omega _{j,1}=0
\end{equation}
and
\begin{equation}
\omega _{j,0} =  \pm {\sqrt {j(N-j)}}.
\label{eq:omegaorigin}
\end{equation}
Without loss of generality (due to (\ref{eq:omegaswap})),
we take the positive square root in (\ref{eq:omegaorigin}).

The expansions (\ref{eq:origin1}) are substituted into the field equations (\ref{eq:YMe},\ref{eq:Ee}) to
determine the values of the remaining constants in the expansions.
The details of this analysis are presented in Section \ref{sec:origin}, but the upshot
is as follows.
We define a vector $\bomega (r)=\left( \omega _{1}(r), \omega _{2}(r), \ldots ,
\omega _{N-1}(r) \right) ^{T}$.
It turns out that, in order to have $N-1$ independent parameters to describe the $N-1$ gauge field
degrees of freedom $\omega _{j}(r)$, it is necessary to expand the $\omega _{j}(r)$ to order $O(r^{N})$.
Then the expansions (\ref{eq:origin1}) become
\begin{eqnarray}
m(r) & = & m_{3}r^{3} + O(r^{4});
\nonumber \\
S(r) & = & S_{0} + O(r^{2});
\nonumber \\
\bomega (r) & = & \bomega _{0} +
\sum _{k=2}^{N} z_{k}(r) {\bmath {v}}_{k} r^{k} + O(r^{N+1}),
\label{eq:origin}
\end{eqnarray}
where
\begin{equation}
\bomega _{0} = \left( {\sqrt {N-1}}, {\sqrt {2(N-2)}}, \ldots , {\sqrt {(N-1)}} \right) ^{T}.
\end{equation}
The functions $z_{k}(r)$, $k=2, \ldots ,N$ are determined completely by the arbitrary constants
${\bar {z}}_{k}=z_{k}(0)$ for $k=2,\ldots , N$.
Together with $\Lambda $, the constants ${\bar {z}}_{2},\ldots , {\bar {z}}_{N}$  determine the constant $m_{3}$
(see equation (\ref{eq:mk+1})).
The $N-1$ vectors ${\bmath {v}}_{k}$ are constant and their components are fixed (see equation (\ref{eq:vcomponents})).
The expansions (\ref{eq:origin}) therefore
give the field variables in terms of the $N+1$ parameters ${\bar {z}}_{2},\ldots , {\bar {z}}_{N}, \Lambda $
and $S_{0}$.

\subsubsection{Event horizon}
\label{sec:bchorizon}

For black hole solutions, we assume that there is a regular, non-extremal event horizon
at $r=r_{h}$, where $\mu (r)$ has a single zero.
This fixes the value of $m(r_{h})$ to be
\begin{equation}
2m( r_{h} ) = r_{h} - \frac {\Lambda r_{h}^{3}}{3}.
\end{equation}
We assume that the field variables $\omega _{j}(r)$, $m(r)$ and $S(r)$ have regular Taylor series
expansions about $r=r_{h}$:
\begin{eqnarray}
m(r) & = & m (r_{h}) + m' (r_{h}) \left( r - r_{h} \right)
+ O \left( r- r_{h} \right) ^{2} ;
\nonumber \\
\omega _{j} (r) & = & \omega _{j}(r_{h}) + \omega _{j}' (r_{h}) \left( r - r_{h} \right)
+ O \left( r -r_{h} \right) ^{2};
\nonumber \\
S(r) & = & S(r_{h}) + S'(r_{h}) \left( r-r_{h} \right)
+ O\left( r - r_{h} \right) .
\label{eq:horizon}
\end{eqnarray}
Setting $\mu (r_{h})=0$ in the Yang-Mills equations (\ref{eq:YMe}) fixes the derivatives of the
gauge field functions at the horizon:
\begin{equation}
\omega _{j} ' (r_{h}) = - \frac {W_{j}(r_{h})\omega _{j}(r_{h})}{2m(r_{h}) - 2r_{h}^{3} p_{\theta } (r_{h})
-\frac {2\Lambda r_{h}^{3}}{3}}.
\end{equation}
Therefore the expansions (\ref{eq:horizon}) are determined by
the $N+1$ quantities $\omega _{j}(r_{h})$, $r_{h}$, $S(r_{h})$ for fixed
cosmological constant $\Lambda $.
For the event horizon to be non-extremal, it must be the case that
\begin{equation}
2m'(r_{h}) = 2r_{h}^{2} p_{\theta} (r_{h}) < 1- \Lambda r_{h}^{2},
\label{eq:constraint}
\end{equation}
which weakly constrains the possible values of the gauge field functions $\omega _{j}(r_{h}) $
at the event horizon.
Since the field equations (\ref{eq:YMe},\ref{eq:Ee}) are invariant under the transformation
(\ref{eq:omegaswap}), we may consider
$\omega _{j}(r_{h}) >0$ without loss of generality.

\subsubsection{Infinity}
\label{sec:bcinfinity}

At infinity, we require that the metric (\ref{eq:metric}) approaches adS, and therefore
the field variables $\omega _{j}(r)$, $m(r)$ and $S(r)$ converge to constant values as
$r\rightarrow \infty $.
We assume that the field variables have regular Taylor series expansions in $r^{-1}$ near infinity:
\begin{eqnarray}
m(r)  & = &  M + O \left( r^{-1} \right) ;
\qquad
S(r) = 1 + O\left( r^{-1} \right) ;
\nonumber \\
\omega _{j}(r) & = & \omega _{j,\infty } + O \left( r^{-1} \right) .
\label{eq:infinity}
\end{eqnarray}
If the space-time is asymptotically flat, with $\Lambda =0$, then the values of $\omega _{j,\infty }$
are constrained to be
\begin{equation}
\omega _{j,\infty } = \pm {\sqrt {j(N-j)}} .
\label{eq:AFinfinity}
\end{equation}
This condition means that the asymptotically flat black holes have no magnetic charge at infinity, or,
in other words, these solutions have no global magnetic charge.
Therefore, at infinity, they are indistinguishable from Schwarzschild black holes.
However, if the cosmological constant is negative,
then there are no {\it {a priori}} constraints on the values of $\omega _{j,\infty }$.
In general, therefore, the adS black holes will be magnetically charged.
The fact that the boundary conditions at infinity in the $\Lambda <0$ case
are less restrictive than those in the $\Lambda = 0$ case
leads to the expectation of many more solutions when $\Lambda <0$ compared with $\Lambda =0$.

\subsection{Embedded solutions}
\label{sec:su2embedded}

The field equations (\ref{eq:YMe},\ref{eq:Ee}) are non-linear and coupled, but they do have two analytic,
trivial solutions.
\begin{description}
\item[Schwarzschild-adS]
Setting
\begin{equation}
\omega _{j}(r) \equiv \pm {\sqrt {j(N-j)}}
\label{eq:SadS}
\end{equation}
for all $j$ gives the Schwarzschild-adS black hole
with
\begin{equation}
m(r) =M= {\mbox {constant}}
\end{equation}
We note that, by setting $M=0$, pure adS space is also a solution.
\item[Reissner-Nordstr\"om-adS]
Setting
\begin{equation}
\omega _{j}(r) \equiv 0
\label{eq:RNadS}
\end{equation}
for all $j$ gives the Reissner-Nordstr\"om-adS black hole
with metric function
\begin{equation}
\mu (r) = 1 - \frac {2M}{r} + \frac {Q^{2}}{r^{2}} - \frac {\Lambda r^{2}}{3},
\end{equation}
where the magnetic charge $Q$ is fixed by
\begin{equation}
Q^{2} = \frac {1}{6} N \left( N+1 \right) \left( N-1 \right) .
\end{equation}
Only for this value of the magnetic charge is the Reissner-Nordstr\"om-adS black hole a solution
of the field equations.
\end{description}

As well as these effectively abelian embedded solutions, there is also a class of embedded ${\mathfrak {su}}(2)$
non-abelian solutions.
These will turn out to play an important role in our proof of the existence of genuinely ${\mathfrak {su}}(N)$
solutions in Section \ref{sec:existence}.
The construction of the embedded ${\mathfrak {su}}(2)$ solutions in adS follows from that given in \cite{Kunzle2} in
the asymptotically flat case.
First, we write the $N-1$ gauge field functions $\omega _{j}(r)$ in terms of a single function $\omega (r)$ as follows:
\begin{equation}
\omega _{j}(r) =\pm  {\sqrt {j(N-j)}} \, \omega (r) \qquad \forall j=1,\ldots ,N-1,
\label{eq:embeddedsu2}
\end{equation}
then define \cite{Kunzle2}
\begin{equation}
\lambda _{N} = {\sqrt {\frac {1}{6} N \left( N -1 \right) \left( N +1  \right) }}.
\label{eq:lambdaN}
\end{equation}
Next we rescale the field variables as follows:
\begin{eqnarray}
R & = &  \lambda _{N}^{-1} r; \qquad
{\tilde {\Lambda }} = \lambda _{N}^{2}\Lambda ; \qquad
{\tilde {m}}(R)  =  \lambda _{N}^{-1} m(r);
\nonumber \\
{\tilde {S}}(R) & =  & S(r); \qquad
{\tilde {\omega }}(R)  =  \omega (r).
\label{eq:scaling}
\end{eqnarray}
Note that we rescale the cosmological constant $\Lambda $ (this is not necessary in \cite{Kunzle2} as
there $\Lambda = 0$).
The field equations satisfied by ${\tilde {m}}(R)$, ${\tilde {S}}(R)$ and ${\tilde {\omega }}(R)$ are then
\begin{eqnarray}
\frac {d{\tilde {m}}}{dR} & = &
 \mu {\tilde {G}} + R^{2} {\tilde {p}}_{\theta }  ; \qquad
\nonumber \\
\frac {1}{{\tilde {S}}} \frac {d{\tilde {S}}}{dR} & = &
- \frac {2{\tilde {G}}}{R} ;
\nonumber \\
0 & = & R^{2} \mu \frac {d^{2}{\tilde {\omega }}}{dR^{2}} +
\left[ 2{\tilde {m}} - 2R^{3} {\tilde {p}}_{\theta } - \frac {2{\tilde {\Lambda }}R^{3}}{3}
\right] \frac {d{\tilde {\omega }}}{dR}
+ \left[ 1 - {\tilde {\omega }}^{2} \right] {\tilde {\omega }};
\label{eq:su2equations}
\end{eqnarray}
where we now have
\begin{equation}
\mu  =  1 - \frac {2{\tilde {m}}}{R} - \frac {{\tilde {\Lambda }}R^{2}}{3},
\end{equation}
and
\begin{equation}
{\tilde {G}} = \left( \frac {d{\tilde {\omega }}}{dR} \right) ^{2} , \qquad
{\tilde {p}}_{\theta } = \frac {1}{2R^{4}} \left( 1 - {\tilde {\omega }}^{2} \right) ^{2}.
\end{equation}
The equations (\ref{eq:su2equations}) are precisely the ${\mathfrak {su}}(2)$ EYM field equations.
Furthermore, the boundary conditions (\ref{eq:origin},\ref{eq:horizon},\ref{eq:infinity}) also reduce to
those for the ${\mathfrak {su}}(2)$ case.
This is straightforward to see for the boundary conditions at the horizon (\ref{eq:horizon}) or
at infinity (\ref{eq:infinity}).
At the origin, the ${\mathfrak {su}}(2)$ embedded solutions are given by ${\bar {z}}_{2}\neq 0$, with
${\bar {z}}_{3}={\bar {z}}_{4} = \ldots = {\bar {z}}_{N}=0$.
Therefore any ${\mathfrak {su}}(2)$, asymptotically adS, EYM soliton or black hole solution can be embedded
into ${\mathfrak {su}}(N)$ EYM to give an asymptotically adS soliton or black hole.
Given that the solution space of ${\mathfrak {su}}(2)$ EYM solitons and black holes in adS has been extensively
studied \cite{Winstanley1,Breitenlohner1,Baxter2}, we immediately have much useful information about the space of solutions of
${\mathfrak {su}}(N)$ EYM soltions and black holes.
However, our purpose in this paper is to prove the existence of genuinely ${\mathfrak {su}}(N)$ (that is,
not embedded) solutions.

\subsection{Properties of the numerical ${\mathfrak {su}}(N)$ solutions}
\label{sec:numerics}

The numerical solution of the field equations (\ref{eq:YMe},\ref{eq:Ee}) and the properties of the
space of solutions are discussed in detail in \cite{Baxter2}.
In this section we briefly review the salient features of the solution space for our purposes here,
and illustrate these features with some figures for the ${\mathfrak {su}}(3)$ case.

\begin{figure}[p]
\begin{center}
\includegraphics[width=8cm,angle=270]{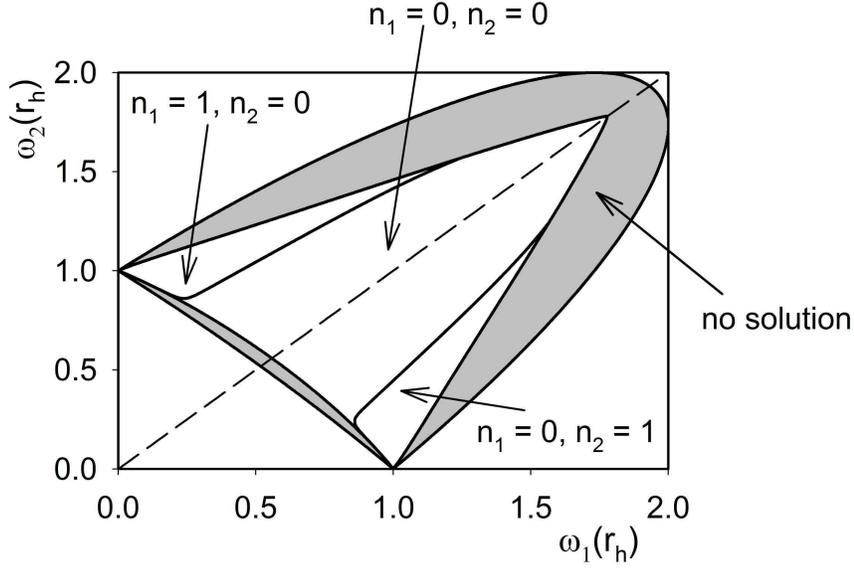}
\end{center}
\caption{Solution space for ${\mathfrak {su}}(3)$ black holes with $r_{h}=1$ and $\Lambda = -2$.
The numbers of zeros of the gauge field functions, $n_{1}$ and $n_{2}$, are shown for the various
regions of the solution space.
The grey region denotes those areas of the $\left( \omega _{1}(r_{h}), \omega _{2}(r_{h}) \right) $
plane for which the constraint (\ref{eq:constraint}) is satisfied, but for which we do not find solutions.
Outside the shaded region the constraint (\ref{eq:constraint}) is not satisfied.
The dashed line is $\omega _{1}(r_{h})=\omega _{2}(r_{h})$, along which lie the embedded ${\mathfrak {su}}(2)$
solutions.
}
\label{fig:su3bhlambda-2}
\end{figure}
\begin{figure}[p]
\begin{center}
\includegraphics[width=8cm,angle=270]{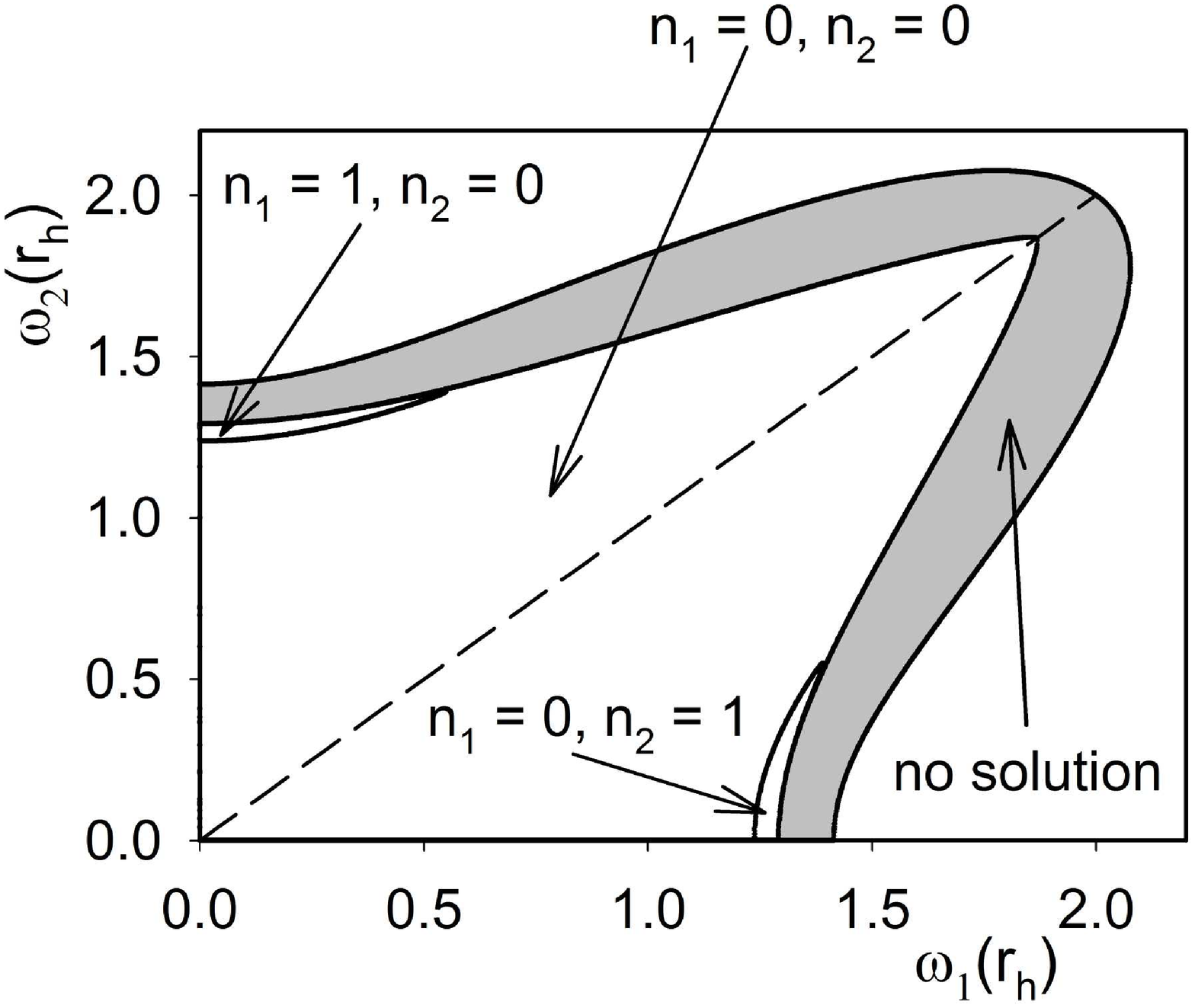}
\end{center}
\caption{Solution space for ${\mathfrak {su}}(3)$ black holes with $r_{h}=1$ and $\Lambda = -3$.
The numbers of zeros of the gauge field functions, $n_{1}$ and $n_{2}$, are shown for the various
regions of the solution space.
The grey region denotes those areas of the $\left( \omega _{1}(r_{h}), \omega _{2}(r_{h}) \right) $
plane for which the constraint (\ref{eq:constraint}) is satisfied, but for which we do not find solutions.
Outside the shaded region the constraint (\ref{eq:constraint}) is not satisfied.
The dashed line is $\omega _{1}(r_{h})=\omega _{2}(r_{h})$, along which lie the embedded ${\mathfrak {su}}(2)$
solutions.
}
\label{fig:su3bhlambda-3}
\end{figure}
\begin{figure}[p]
\begin{center}
\includegraphics[width=8cm,angle=270]{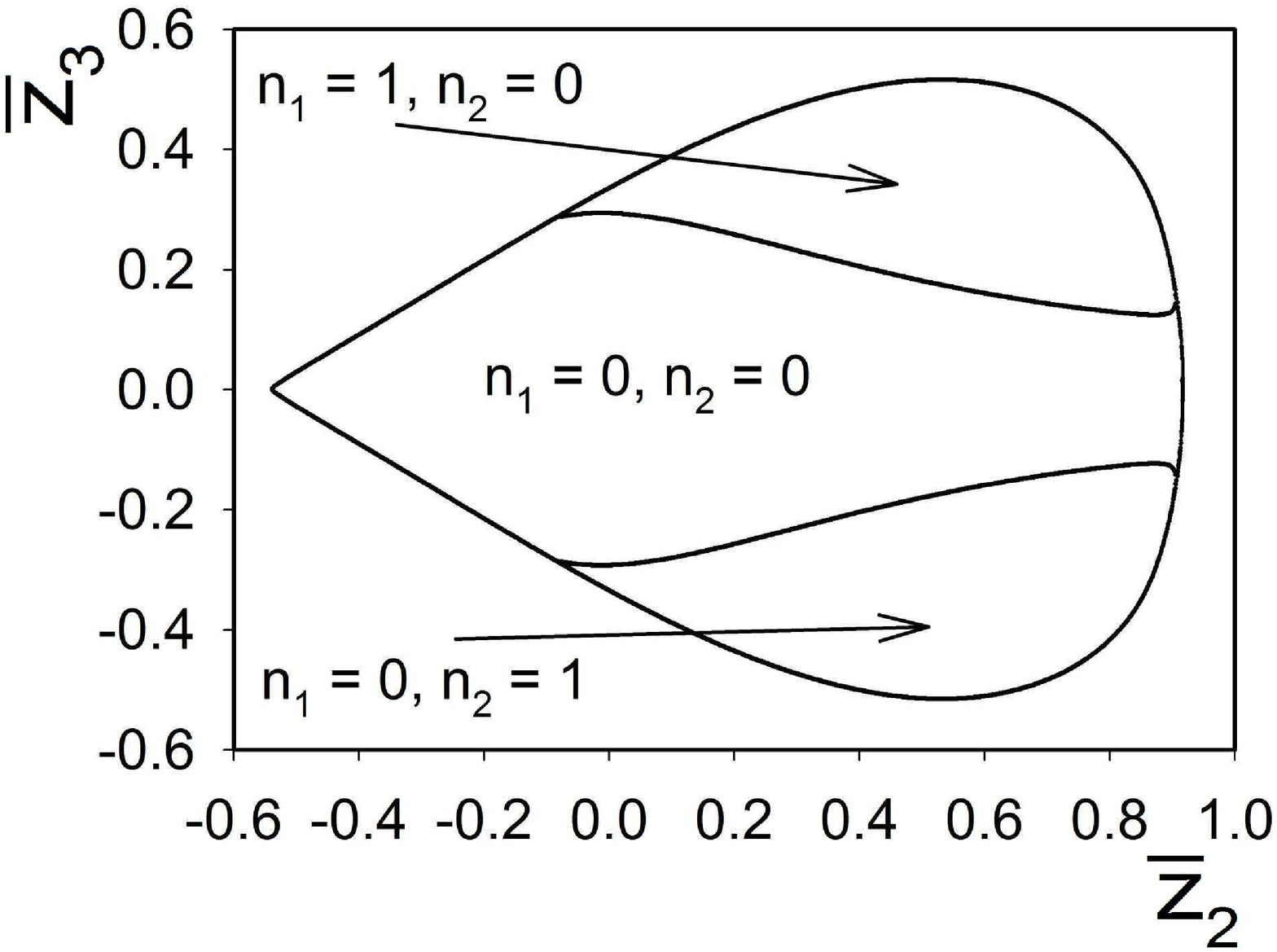}
\end{center}
\caption{Solution space for ${\mathfrak {su}}(3)$ solitons with $\Lambda = -2$.
The numbers of zeros of the gauge field functions, $n_{1}$ and $n_{2}$, are shown for the various
regions of the solution space.
}
\label{fig:su3solitonlambda-2}
\end{figure}
\begin{figure}[p]
\begin{center}
\includegraphics[width=8cm,angle=270]{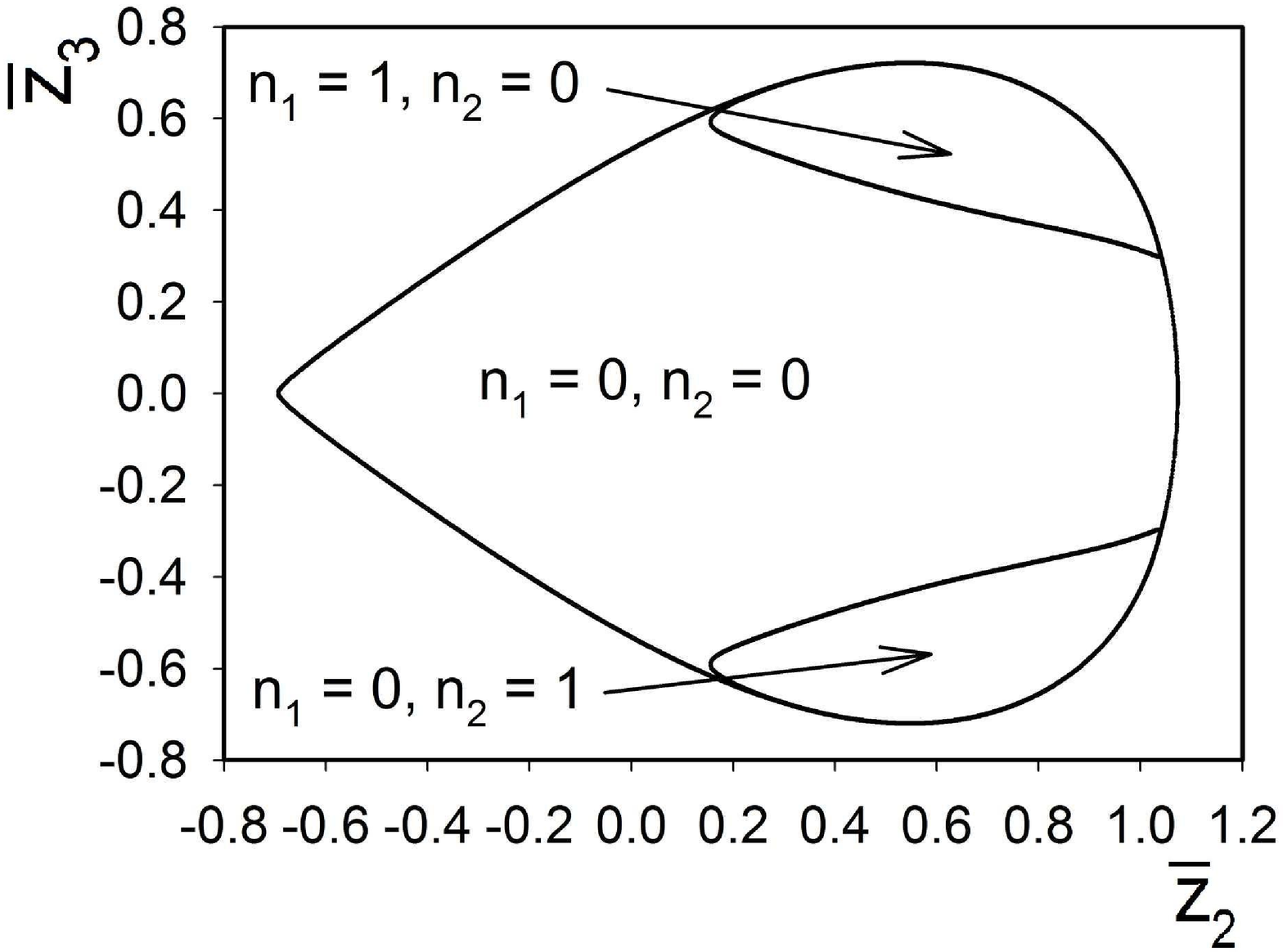}
\end{center}
\caption{Solution space for ${\mathfrak {su}}(3)$ black holes with $\Lambda = -3$.
The numbers of zeros of the gauge field functions, $n_{1}$ and $n_{2}$, are shown for the various
regions of the solution space.
}
\label{fig:su3solitonlambda-3}
\end{figure}

As in the ${\mathfrak {su}}(2)$ case \cite{Winstanley1,Bjoraker}, we find solutions
in open sets of the parameter space
( $\left[ {\bar {z}}_{2}, \ldots {\bar {z}}_{N} \right] $ (\ref{eq:origin}) for solitons and
$\left[ \omega _{1}(r_{h}),\ldots \omega _{N-1}(r_{h}) \right] $ for black holes).
For all negative values of the cosmological constant $\Lambda <0$, we find solutions in which
all the gauge field functions have no zeros, in a neighbourhood of the embedded pure adS (soliton)
or Schwarzschild-adS (black hole) solution, although in practice the size of this neighbourhood
decreases rapidly as $\left| \Lambda \right| $ decreases.
The existence of such a region of solutions where all gauge field functions have no zeros is proven
in Section \ref{sec:result}, see Corollary \ref{thm:corollary}.
As $\left| \Lambda \right| $ decreases, the region of phase space where we have soliton or
black hole solutions shrinks and breaks up, eventually becoming discrete points in
the limit $\left| \Lambda \right| \rightarrow 0$.

The main interest in this paper is solutions for which the gauge field functions have no zeros,
and particularly the behaviour of the space of solutions for large $\left| \Lambda \right| $.
As $\left| \Lambda \right| $ increases, the region of phase space where solutions exist expands rapidly,
as can be seen by comparing the size of the regions for $\Lambda = -2$ and $\Lambda = -3$ in
Figures \ref{fig:su3bhlambda-2} and \ref{fig:su3bhlambda-3} (for black holes)
and Figures \ref{fig:su3solitonlambda-2} and \ref{fig:su3solitonlambda-3} (for solitons).
It can also be seen in these figures that the regions of solutions where at least one gauge field function
has at least one zero shrink as $\left| \Lambda \right| $ increases.
For black holes, Figure 10 in \cite{Baxter2} shows that there are no black hole solutions for which at least one gauge
field function has at least one zero when $\Lambda = -5$.
As $\left| \Lambda \right| $ increases still further, all the solutions which exist are such that all the gauge field
functions have no zeros.
In addition, the region of phase space where there are solutions continues to expand without limit as $\left|
\Lambda \right| $ increases.
We will show in Section \ref{sec:largeL} that for any point in the phase space, for all sufficiently large
$\left| \Lambda \right| $, this point leads to a black hole or soliton solution (as applicable) for which
all the gauge field functions have no zeros.

We have ${\mathfrak {su}}(2)$ embedded solutions (\ref{eq:embeddedsu2}) along the lines
$\omega _{1}(r_{h})=\omega _{2}(r_{h})$ for black hole solutions (indicated by a dashed line in
Figures \ref{fig:su3bhlambda-2}--\ref{fig:su3bhlambda-3}) and ${\bar {z}}_{3}=0$ for soliton solutions
(not everywhere along the lines, only where the lines are contained within the region of solutions).
The existence of genuinely ${\mathfrak {su}}(N)$ solutions in a neighbourhood of embedded ${\mathfrak {su}}(2)$
solutions, seen in Figures \ref{fig:su3bhlambda-2}--\ref{fig:su3solitonlambda-3},
is shown in Section \ref{sec:result}.

\section{Local existence of solutions}
\label{sec:local}

The first part of our proof of the existence of regular soliton and black hole
solutions of the ${\mathfrak {su}}(N)$ EYM equations (\ref{eq:YMe},\ref{eq:Ee})
is showing that regular solutions exist in neighbourhoods of the
points $r=0$, $r=r_{h}$ and $r=\infty $, where the field equations
are singular.

In each case, our method follows that employed in \cite{Breitenlohner2}
for asymptotically flat ${\mathfrak {su}}(2)$ EYM, and subsequently
generalized in asymptotically flat space to any compact gauge group in \cite{Oliynyk1}.
For ${\mathfrak {su}}(N)$ EYM in asymptotically flat space, the paper \cite{Kunzle2} used
a different approach to prove local existence.

The method we employ here makes use of the following theorem, which is taken from \cite{Breitenlohner2}, but
which we restate for convenience:
\begin{theorem}
\label{thm:BFM}
\cite{Breitenlohner2} Consider a system of differential equations for $n+m$ functions
${\bmath {u}}=\left( u_{1},u_{2},\ldots u_{n} \right) ^{T}$
and ${\bmath {v}} = \left( v_{1}, v_{2}, \ldots v_{m} \right) ^{T}$
of the form
\begin{eqnarray}
x \frac {du_{i}}{dx} = x^{\sigma _{i}} f_{i}\left(x,{\bmath {u}},{\bmath {v}}\right) ;
\nonumber \\
x \frac {dv_{i}}{dx} = - \tau _{i} v_{i} + x^{\varsigma _{i}} g_{i} \left( x, {\bmath {u}}, {\bmath {v}} \right) ;
\label{eq:BFMsys}
\end{eqnarray}
with constants $\tau _{i}>0$ and integers $\sigma _{i}, \varsigma _{i} \ge 1$, and let ${\cal {C}}$ be
an open subset of ${\mathbb {R}}^{n}$ such that the functions $f_{i}$, $i=1,\ldots n$ and $g_{i}$, $i=1,\ldots m$
are analytic in a neighbourhood of $x=0$, ${\bmath {u}}={\bmath {c}}$, ${\bmath {v}}={\bmath {0}}$
for all ${\bmath {c}} \in {\cal {C}}$.
Then there exists an $n$-parameter family of solutions of the system (\ref{eq:BFMsys})
such that
\begin{equation}
u_{i}(x) = c_{i} + O\left( x^{\sigma _{i}} \right) ,
\qquad
v_{i}(x) = O \left( x^{\varsigma _{i}} \right) ,
\end{equation}
where $u_{i}(x)$, $i=1,\ldots n$ and $v_{i}(x)$, $i=1,\ldots m$ are defined for ${\bmath {c}} \in {\cal {C}}$,
for $\left| x \right| < x_{0}({\bmath {c}})$ (for some $x_{0}({\bmath {c}})>0)$ and are analytic
in $x$ and ${\bmath {c}}$.
\end{theorem}
This theorem allows one to parameterize the family of solutions near a singular point of a set of ordinary
differential equations.
The key part of the method is putting the field equations into the form (\ref{eq:BFMsys}) with $x=0$ at
the singular point, which requires a change of variables.
We need to consider the three singular points $r=0$, $r=r_{h}$ and $r\rightarrow \infty $ separately.
Once we have the field equations in the relevant form, it is then straightforward to prove local existence
of solutions of the field equations with the required behaviour (\ref{eq:origin},\ref{eq:horizon},\ref{eq:infinity}).
The proof of the local existence of solutions near the origin is by far the most complicated of the three.
The fact that the local solutions are analytic in the parameters $c_{i}$ as well as the independent variable
$x$ will turn out to be extremely useful in Section \ref{sec:existence}.

\subsection{Local existence of solutions near the origin}
\label{sec:origin}

The local existence of solutions of the field equations near the origin is technically rather complicated.
Local existence near the origin has been proved in the asymptotically flat case for ${\mathfrak {su}}(N)$
gauge fields in \cite{Kunzle2}, using a method which is different from that employed here, and
for general compact gauge groups in \cite{Oliynyk1}, whose method we follow.
As might be expected, the inclusion of a negative cosmological constant $\Lambda $ does not significantly change the
analysis.
However, the analysis is highly involved and in particular it is important to verify exactly how the cosmological
constant appears in the derivation.
For this reason, in this section we have included sufficient detail to make it clear where
our analysis differs from that in \cite{Oliynyk1,Kunzle2}.

We begin by following \cite{Kunzle2} and define
\begin{equation}
\gamma _{j} = j \left( N - j \right), \qquad j = 0, \ldots , N .
\end{equation}
Then we define new variables $u_{j}(r)$ and $q_{j}(r)$ as follows \cite{Kunzle2}:
\begin{eqnarray}
\omega _{j}(r) & = & u_{j} (r) \gamma _{j}^{\frac {1}{2}}; \qquad j=1, \ldots , N-1;
\nonumber \\
q_{j} (r) & = & \omega _{j}^{2} - \omega _{j-1}^{2} - N - 1 +2j ; \qquad j = 1, \ldots , N.
\end{eqnarray}
Next define a new independent variable $x$ by
\begin{equation}
x = \frac {r}{\lambda _{N} }
\label{eq:xorigin}
\end{equation}
where $\lambda _{N} $ is defined in (\ref{eq:lambdaN}).
We then write the cosmological constant $\Lambda $ and metric function $m(r)$ in terms of new quantities
${\tilde {\Lambda }}$ and ${\tilde {m}}$ as follows:
\begin{equation}
m(r) = \lambda _{N} {\tilde {m}}(x); \qquad \Lambda = \frac {{\tilde {\Lambda }}}{\lambda _{N} ^{2}} ;
\end{equation}
and from now on consider $u_{j}(r)=u_{j}(\lambda _{N} x)$ and $S(r)=S(\lambda _{N} x)$ as functions of $x$ instead of $r$.
In terms of these variables the field equations become \cite{Kunzle2}
\begin{eqnarray}
\frac {d{\tilde {m}}}{dx} & = & \frac {1}{\lambda _{N} ^{2}} \left[ \mu {\tilde {G}} + {\tilde {P}} \right] ;
\qquad
\frac {dS}{dx} = \frac {2S{\tilde {G}}}{\lambda _{N} ^{2}x};
\label{eq:scaledequation1}
 \\
0 & = &
x^{2} \mu \frac {d^{2}u_{j}}{dx^{2}} + \left[ 2{\tilde {m}} - \frac {2{\tilde {\Lambda }}x^{3}}{3}
- \frac {2x}{\lambda _{N} ^{2}}{\tilde {P}} \right] \frac {du_{j}}{dx}
+ \frac {1}{2} \left[ q_{j+1}-q_{j} \right] u_{j},
\label{eq:scaledequation2}
\end{eqnarray}
where
\begin{equation}
{\tilde {G}}  =
\sum _{j=1}^{N-1} \gamma _{j} \left( \frac {du_{j}}{dx} \right) ^{2};
\qquad
{\tilde {P}}  = \frac {1}{4x^{2}} \sum _{j=1}^{N} q_{j}^{2}.
\label{eq:tildeG&P}
\end{equation}
The remainder of our analysis of the solutions of the field equations near the origin is split into two parts.
Firstly, in Section \ref{sec:powerseries}, we derive the power series expansion (\ref{eq:origin}) for the field
variables near the origin.
Secondly, in Section \ref{sec:originproof}, we prove the local existence of solutions of the field equations having
this form near the origin.
In this section, we will not be using the Einstein summation convention, and all summations are written
explicitly.

\subsubsection{Derivation of the power series expansion}
\label{sec:powerseries}

In this section we seek a power series expansion for ${\tilde {m}}$ and $u_{j}$ in a neighbourhood of the origin
$x=0$, in the form
\begin{equation}
{\tilde {m}}(x) = \sum _{k=0}^{\infty } m_{k}x^{k}; \qquad
u_{j} = \sum _{k=0}^{\infty } u_{j,k} x^{k} .
\end{equation}
A straightforward analysis of the field equations shows that $m_{0}$, $m_{1}$, $m_{2}$ and $u_{j,1}$ are all zero
and that $u_{j,0}=1$.

We seek to substitute these power series expansions into the scaled field equations
(\ref{eq:scaledequation1},\ref{eq:scaledequation2}).
First we require the power series expansions for the quantities $q_{j}$, ${\tilde {G}}$ and ${\tilde {P}}$
in terms of the $m_{k}$ and $u_{j,k}$.
These do not depend on ${\tilde {\Lambda }}$ and so have the same form as in \cite{Kunzle2}:
\begin{equation}
q_{j} = \sum _{k=2}^{\infty } q_{j,k} x^{k}; \qquad
{\tilde {G}} = \sum _{k=2}^{\infty } G_{k} x^{k}; \qquad
{\tilde {P}} = \sum _{k=2}^{\infty } P_{k} x^{k};
\end{equation}
where
\begin{eqnarray}
q_{j,k} & = &
\sum _{\ell =0}^{k} \gamma _{j} u_{j,\ell } u_{j, k-\ell } - \gamma _{j-1} u_{j-1,\ell } u_{j-1, k-\ell } ;
\nonumber \\
 G_{k} & = &
 \sum _{j=1}^{N-1}  \sum _{\ell =1}^{k-1} \gamma _{j}
 \left( \ell + 1 \right) \left( k- \ell + 1 \right) u_{j,\ell +1} u_{j,k-\ell + 1}  ;
 \nonumber \\
 P_{k} & = & \frac {1}{4} \sum _{j=1}^{N} \sum _{\ell =2}^{k} q_{j, \ell} q_{j,k-\ell +2 }.
\end{eqnarray}
It is also helpful to have a power series expansion for $\mu $:
\begin{equation}
\mu = \sum _{k=0}^{\infty } \mu _{k}x^{k},
\end{equation}
where
\begin{equation}
\mu _{0}=1, \qquad \mu _{1}=0, \qquad
\mu _{2} = -2 m_{3} - \frac {{\tilde {\Lambda }}}{3};
\end{equation}
and, for $k\ge 3$,
\begin{equation}
\mu _{k} = -2m_{k+1}.
\end{equation}
It is clear that there is a minor difference in this expansion compared with the asymptotically flat case, due
to the dependence of $\mu _{2}$ on ${\tilde {\Lambda }}$.
It is this small difference of which we must carefully keep track
if we are to verify that the analysis of \cite{Kunzle2}
holds when ${\tilde {\Lambda }} \neq 0$.

Examining the coefficient of $x^{k}$ in the first equation (\ref{eq:scaledequation1}), we find, for $k\ge 2$:
\begin{equation}
m_{k+1} = \frac {1}{\lambda _{N} ^{2} \left( k + 1 \right) } \left[
P_{k} + \sum _{\ell = 0}^{k-2} \mu _{\ell }G_{k-\ell } \right] .
\label{eq:mk+1}
\end{equation}
Here, $P_{k}$ depends on $q_{j,2},\ldots , q_{j,k}$ and hence on $u_{j,0},\ldots ,u_{j,k}$;
the quantity $\mu _{\ell }$ $(\ell = 0, \ldots , k-2 )$ depends on $m_{3},\ldots , m_{k-1}$ and ${\tilde {\Lambda }}$,
and $G_{\ell }$ $(\ell = 2,\ldots , k)$ depends on $u_{j,2},\ldots , u_{j,k}$.
Therefore equation (\ref{eq:mk+1}) determines $m_{k+1}$ uniquely in terms of $m_{3},\ldots , m_{k-1}, u_{j,0},
\ldots , u_{j,k}$ and ${\tilde {\Lambda }}$.
Apart from the dependence on ${\tilde {\Lambda }}$, this is exactly the same situation as in \cite{Kunzle2}.

Next we turn to the more complicated equation (\ref{eq:scaledequation2}).
After a great deal of algebra, we obtain, from the coefficient of $x^{k}$ for $k\ge 2$, the following equation:
\begin{equation}
b_{j,k} = 2\gamma _{j} u_{j,k} - k \left( k-1 \right) u_{j,k} + \gamma _{j+1} u_{j+1,k}
- \gamma _{j-1} u_{j-1,k} \, ;
\label{eq:bjk}
\end{equation}
for $j=1,\ldots ,N-1$,
where $b_{j,k}$ is a complicated expression which we write as
\begin{equation}
b_{j,k} = c_{j,k} + {\tilde {c}}_{j,k},
\end{equation}
where
\begin{eqnarray}
c_{j,k} & = &
- \sum _{\ell = 2}^{k-2} u_{j,\ell } V_{k, \ell } ;
\nonumber \\
{\tilde {c}}_{j,k} & = &
-\frac {1}{2} \sum _{\ell =2}^{k-2} \sum _{i=1}^{N-1} A_{i,j} u_{i,k} u_{i,k-\ell }
 - \sum _{\ell = 2}^{k-2} \sum _{i=1}^{N-1} A_{i,j} u_{i,k} u_{j,k-\ell }
\nonumber \\ & &
 - \sum _{\ell = 2}^{k-2} \sum _{p=2}^{k-\ell -2} \sum _{i=1}^{N-1} A_{i,j} u_{j,p} u_{j,k-\ell - 2 - p} ;
\end{eqnarray}
and we have defined further quantities $V_{k, \ell }$ and $A_{i,j}$ as
\begin{eqnarray}
V_{k , \ell } & = &
\frac {{\tilde {\Lambda }}}{3} \left( k-2 \right) \left( k-1 \right) \delta _{\ell , k-2 }
+ \frac {2}{\lambda _{N} ^{2}} \ell P_{k- \ell } + 2\ell \left( \ell -2 \right) m_{k-\ell + 1} ;
\nonumber \\
A_{i,j} & = &
\gamma _{j} \left[ 2\delta _{i,j} - \delta _{i+1,j} - \delta _{i-1,j} \right] ;
\label{eq:Adef}
\end{eqnarray}
and $\delta _{i,j}$ is the usual Kronecker $\delta $.
Using the $(N-1)\times(N-1)$ matrix $A$ whose entries are $A_{i,j}$ we can write equation (\ref{eq:bjk}) as
\begin{equation}
b_{j,k} = \sum _{i=1}^{N-1} \left[ A_{i,j} - \delta _{i,j} k \left( k-1 \right) \right] u_{i,k} .
\label{eq:brecurrence}
\end{equation}
This is exactly the same as the equation for the corresponding quantities $b_{j,k}$ in the asymptotically
flat case \cite{Kunzle2}.
However, our $b_{j,k}$ is slightly different from that in \cite{Kunzle2}, as $V_{j,k}$ contains ${\tilde {\Lambda }}$.
We have examined the analysis in \cite{Kunzle2} very carefully, and found that
$V_{j,k}$ arises only in Lemma 3 in \cite{Kunzle2}, and, even in the proof of Lemma 3, the precise form of
$V_{j,k}$ is not important.
Therefore Theorem 1 in \cite{Kunzle2} holds also in our case, and we restate it here for convenience:
\begin{theorem}
\cite{Kunzle2}
The recurrence relations (\ref{eq:mk+1}) and (\ref{eq:brecurrence}) determine uniquely all the coefficients $m_{k}$
and $u_{j,k}$ for $k>N$ once $N-1$ arbitrary parameters have been chosen, one for each equation with $k=2, \ldots, N$.
\label{thm:Kunzle}
\end{theorem}
We comment that in this section we have taken $k=2,\ldots N$, whereas in \cite{Kunzle2}, the parameter $k$ has
values $k=1,\ldots N-1$, and the quantity $u_{j,k}$ multiplies $x^{k+1}$ in \cite{Kunzle2} rather than $x^{k}$ here.
Some care is therefore needed in comparing results in \cite{Kunzle2} with those in this section.

The form of the coefficients $u_{j,k}$ for $j=1,\ldots , N-1$,
$k=2,\ldots , N$ is derived in exactly the same way as in \cite{Kunzle2}.
The $(N-1)\times (N-1)$
matrix $A$ (\ref{eq:Adef}) has eigenvalues $k\left( k -1 \right)$ for $k=2,3,\ldots , N$  (proved in \cite{Kunzle2})
with right-eigenvectors ${\bar {{\bmath {v}}}}_{k}$ and left-eigenvectors
${\bar {\bsigma }}_{k}^{T}$, and we write the components of the
vectors ${\bar {{\bmath {v}}}}_{k}$ as ${\bar {v}}_{k,j}$ for $j=1,\ldots , N-1$.
The components $v_{k,j}$ are given in terms of Hahn polynomials \cite{Kunzle2,Hahn}:
\begin{equation}
{\bar {v}}_{k,j} = \frac {N-1}{N-j} {}_{3}F_{2}\left( -k+1 , -j+1, k; 2, -N+1; 1\right) ,
\label{eq:vcomponents}
\end{equation}
where the ${}_{3}F_{2}$ are hypergeometric functions.
We find \cite{Kunzle2}
\begin{equation}
u_{j,2} = {\bar {\beta }}_{2} {\bar {v}}_{2,j} , \qquad
u_{j,3} = {\bar {\beta }}_{3} {\bar {v}}_{3,j}
\label{eq:u2}
\end{equation}
and, for $k=4,\ldots , N$,
\begin{equation}
u_{j,k} = u_{j,k}^{*} + {\bar {\beta }}_{k} {\bar {v}}_{k,j} .
\label{eq:ujk*}
\end{equation}
Here the ${\bar {\beta }}_{k}$, for $k=2,\ldots , N$, are arbitrary constant parameters, and $u_{j,k}^{*}$ is a special
solution of (\ref{eq:brecurrence}) fixed by the requirement that
\begin{equation}
{\bar {\bsigma }}_{k}^{T} {\bmath {u}}_{,k}
= d_{k} {\bar {\beta }}_{k}
\label{eq:ujk*def}
\end{equation}
for $k=2,\ldots , N$, where \cite{Kunzle2}
\begin{equation}
d_{k} = {\bar {\bsigma }}_{k}^{T} {\bar {{\bmath {v}}}}_{k} =
\frac {\left( N+k-1\right) ! \left( N - k \right) !}{\left( N-1 \right) ! \left( N -2 \right) ! k \left( k - 1 \right)
\left( 2k- 1 \right) } ,
\end{equation}
and ${\bmath {u}}_{,k}$ is the vector with components $u_{1,k},\ldots , u_{N-1,k}$.
The $u_{j,k}^{*}$ (\ref{eq:ujk*}) are uniquely determined by Theorem \ref{thm:Kunzle},
and the $N-1$ arbitrary constants ${\bar {\beta }}_{k}$ completely fix the coefficients $u_{j,k}$ (\ref{eq:ujk*}).

It is convenient to expand the vector ${\bmath {u}}_{,k}$ in terms of the eigenvectors of the matrix $A$,
by expanding the particular solutions $u_{j,k}^{*}$ in terms of the ${\bar {{\bmath {v}}}}_{i}$ and collecting terms:
\begin{equation}
{\bmath {u}}_{,k} = \sum _{\ell = 2}^{N} Y_{k}^{\ell } {\bar {{\bmath {v}}}}_{\ell }, \qquad
k=2, \ldots , N.
\end{equation}
The analysis of \cite{Kunzle2} shows that, independent of the form of $V_{j,k}$ (\ref{eq:Adef}),
\begin{equation}
Y_{k}^{\ell } = 0 \qquad {\mbox {for $\ell >k$}},
\end{equation}
so that ${\bmath {u}}_{,k}$ depends only on ${\bar {{\bmath {v}}}}_{2}, \ldots , {\bar {{\bmath {v}}}}_{k}$.
This result is not trivial because it depends on the properties of the particular solutions $u_{j,k}^{*}$.
Furthermore, from \cite{Kunzle2}, the particular solutions $u_{j,k}^{*}$ depend only on the components of the
vectors ${\bar {\bmath {v}}}_{i}$ for $i=2,\ldots , k-1$, so that
\begin{equation}
Y_{k}^{k} = {\bar {\beta }}_{k}.
\end{equation}
Therefore the first $N+1$ terms in the power series for the gauge field functions
${\bmath {u}} = \left( u_{1}, \ldots , u_{N-1} \right) ^{T}$
can be written as
\begin{equation}
{\bmath {u}} = {\bmath {u}}_{0} + \sum _{k=2}^{N} \sum _{\ell =2}^{k} Y _{k}^{\ell } {\bar {{\bmath {v}}}}_{\ell } x^{k}
= {\bmath {u}}_{0} + \sum _{k=2}^{N} \bbeta _{k}(x) x^{k};
\label{eq:ubetaexpansion}
\end{equation}
where ${\bmath {u}}_{0}=\left( 1, \ldots , 1 \right) ^{T}$;
the $Y_{k}^{\ell }$ for $\ell <k$ are determined by ${\bar {\beta }}_{i}$ for $i=2,\ldots ,k-1$;
and we have defined vector functions $\bbeta _{k}(x)$, $k=2, \ldots , N$, by
\begin{equation}
\bbeta _{k}(x) = \sum _{\ell =k}^{N} Y_{\ell }^{k} x^{\ell -k} {\bar {{\bmath {v}}}}_{k}
= {\bar {{\bmath {v}}}}_{k} \left[
{\bar {\beta }}_{k} + Y_{k+1}^{k} x + \ldots + Y_{k+N}^{k} x^{N-k}
\right] .
\label{eq:betavecdef}
\end{equation}
The vectors $\bbeta _{k}(x)$ are proportional to ${\bar {{\bmath {v}}}}_{k}$ for all values of $x$,
and so are eigenvectors of the matrix $A$ (\ref{eq:Adef}) for every $x$.
The constants ${\bar {\beta }}_{k}$, $k=2,\ldots ,N$ in (\ref{eq:betavecdef})
are the same as those in (\ref{eq:ujk*}), and are arbitrary, but the constants
$Y_{k+1}^{k}, \ldots ,Y_{k+N}^{k}$
are determined by the ${\bar {\beta }}_{2},\ldots {\bar {\beta }}_{N}$ in a complicated way.
Returning to the original variables, we find the power series expansion for the gauge field functions
$\omega _{j}$ has the form (\ref{eq:origin}), with the components of $\bbeta _{k}(x)$ determining the
functions $z_{k}(r)$ in (\ref{eq:origin}) as follows:
\begin{equation}
\beta _{k,j}(x) =  z_{k}(r) {\bar {v}}_{k,j} \lambda _{N}^{k} \gamma _{j}^{-\frac {1}{2}},
\end{equation}
where $\bbeta _{k}(x) = \left( \beta _{k,1} , \ldots \beta _{k,N-1} \right) ^{T}$.

The last remaining piece in the power series expansion of the field variables is to find the expansion for $S$.
First let $S=\exp \delta $, so that (\ref{eq:scaledequation1}) becomes
\begin{equation}
\frac {d\delta }{dx} = \frac {2{\tilde {G}}}{\lambda _{N} ^{2}x}.
\end{equation}
From this equation it is clear that $\delta $ has a power series expansion of the form
\begin{equation}
\delta = \delta _{0} + \frac {2}{\lambda _{N} ^{2}} \sum _{k=2}^{\infty } \frac {G_{k}}{k} x^{k};
\end{equation}
for some arbitrary constant $\delta _{0}$.
This then gives a power series expansion for $S$ of the form (\ref{eq:origin}).

\subsubsection{Local existence}
\label{sec:originproof}

To prove the local existence of solutions of the field equations (\ref{eq:YMe},\ref{eq:Ee})
in a neighbourhood of the origin $r=0$, with the power series expansions
(\ref{eq:origin},\ref{eq:ubetaexpansion},\ref{eq:betavecdef})
derived in the previous sub-section, we follow the analysis of \cite{Oliynyk1} in the asymptotically flat
case, valid for any compact gauge group.
As in the previous sub-section, the inclusion of a negative cosmological constant $\Lambda $ does not
significantly change the proof, but it is nonetheless important to carefully consider the details of each step.

\begin{proposition}
There exists an $N+1$-parameter family of local solutions of the field equations
(\ref{eq:YMe},\ref{eq:Ee}) near $r=0$, satisfying the boundary conditions (\ref{eq:origin}),
and analytic in $\Lambda $, ${\bar {z}}_{k}$, $S_{0}$ and $r$.
\label{thm:origin}
\end{proposition}

\paragraph{{\bf {Proof}}}
We begin with the field equations in the form (\ref{eq:scaledequation1},\ref{eq:scaledequation2}).
Defining a vector ${\bmath {u}}= \left( u_{1},\ldots , u_{N-1} \right) ^{T}$,
we write the YM equation (\ref{eq:scaledequation2}) as a vector equation:
\begin{equation}
x^{2} \mu \frac {d^{2}{\bmath {u}}}{dx^{2}} + \left[ 2{\tilde {m}} - \frac {2{\tilde {\Lambda }}x^{3}}{3}
- \frac {2x}{\lambda _{N} ^{2}} {\tilde {P}} \right] \frac {d{\bmath {u}}}{dx} + \frac {1}{2} {\bmath {\cal {W}}} = 0,
\end{equation}
where the vector ${\bmath {\cal {W}}} = \left( {\cal {W}}_{1} , \ldots , {\cal {W}}_{N-1}\right) $
has components
\begin{equation}
{\cal {W}} _{j} = \left( q_{j+1} - q_{j-1} \right) u_{j}
=2u_{j} - \sum _{i=1}^{N-1} u_{j} A_{j,i} u_{i}^{2}
\label{eq:vectorYM}
\end{equation}
and the $(N-1)\times (N-1)$ matrix $A$ is defined in (\ref{eq:Adef}).

Our approach is to use the eigenvectors of the $(N-1)\times (N-1)$ matrix
 $A$ as a basis for our $N-1$ gauge field functions, instead of using
the functions $u_{1},\ldots ,u_{N-1}$.
From the power series derivation in the previous subsection, we are motivated to
write the vector ${\bmath {u}}(x)$ as
a sum over eigenvectors of the matrix $A$ (cf. (\ref{eq:ubetaexpansion})):
\begin{equation}
{\bmath {u}} = {\bmath {u}}_{0} +  \sum _{k=2}^{N} \bbeta _{k}(x) x^{k};
\end{equation}
where ${\bmath {u}}_{0}= \left( 1, \ldots , 1 \right) ^{T}$ and the $\bbeta _{k}(x)$ are vector
functions of $x$ which satisfy
\begin{equation}
A \bbeta _{k} (x) = k \left( k - 1 \right) \bbeta _{k}(x) \qquad {\mbox {for all $x$.}}
\end{equation}
It should be emphasized that we are making no assumptions about the $\bbeta _{k}(x)$ at this stage.
In particular, we are not assuming that $\bbeta _{k}(x)$ are regular at the origin,
nor that they have the form (\ref{eq:betavecdef}).
Since the $N-1$ vectors ${\bar {{\bmath {v}}}}_{k}$ are eigenvectors of the matrix $A$ corresponding to
distinct eigenvalues, they constitute a basis of the space of $N-1$ vectors.
If we write $\bbeta _{k}(x)= {\tilde {\zeta }}_{k}(x) {\bar {{\bmath {v}}}}_{k}$ for
some scalar function ${\tilde {\zeta }}_{k}(x)$ (which will be a constant factor times the $\zeta _{k}(x)$
functions defined below), then we are using the $N-1$ independent functions ${\tilde {\zeta }}_{k}(x)$,
$k=2,\ldots ,N$, instead of the $N-1$ independent functions $u_{j}(x)$, $j=1,\ldots ,N-1$,
to describe our gauge field.

We now multiply both sides of (\ref{eq:vectorYM}) by the left-eigenvectors ${\bar {\bsigma }}_{k}^{T}$ of the matrix $A$:
\begin{equation}
x^{2} \mu {\bar {\bsigma }}_{k}^{T}  \frac {d^{2}{\bmath {u}}}{dx^{2}} +
\left[ 2{\tilde {m}} - \frac {2{\tilde {\Lambda }}x^{3}}{3}
- \frac {2x}{\lambda _{N} ^{2}} {\tilde {P}} \right]
{\bar {\bsigma }}_{k}^{T}  \frac {d{\bmath {u}}}{dx}
+ \frac {1}{2} {\bar {\bsigma }}_{k}^{T} {\bmath {\cal {W}}} = 0.
\end{equation}
Using the fact that the left- and right-eigenvectors of the matrix $A$ are orthogonal \cite{Kunzle2}
we have
\begin{equation}
{\bar {\bsigma }}_{k_{1}}^{T} \bbeta _{k_{2}} = 0 \qquad {\mbox {if $k_{1}\neq k_{2}$}},
\end{equation}
and defining new functions $\zeta _{k}$ by
\begin{equation}
\zeta _{k} (x) = {\bar {\bsigma }}_{k}^{T} \bbeta _{k} ,
\label{eq:zetakdef}
\end{equation}
we obtain
\begin{eqnarray}
0 & = &
x^{2} \mu \left[ x^{k} \frac {d^{2}\zeta _{k}}{dx^{2}} + 2k x^{k-1} \frac {d\zeta _{k}}{dx}
+ k \left( k - 1 \right) x^{k-2} \zeta _{k} \right]
\nonumber \\ & &
+ \left[ 2{\tilde {m}} - \frac {2{\tilde {\Lambda }}x^{3}}{3} - \frac {2x}{\lambda _{N} ^{2}}{\tilde {P}} \right]
\left[ x^{k} \frac {d\zeta _{k}}{dx} +  k x^{k-1} \zeta _{k} \right]
+\frac {1}{2} {\bar {\bsigma }}_{k}^{T} {\bmath {\cal {W}}}.
\label{eq:psi1}
\end{eqnarray}

The complicated part of the analysis lies in casting ${\bmath {\cal {W}}}$ into
a suitable form.
However, the form of ${\bmath {\cal {W}}}$ depends only on the structure of the ${\mathfrak {su}}(N)$
gauge field, without any reference to the form of the metric or the cosmological constant.
Therefore we may appeal to the analysis of \cite{Oliynyk1}, which is valid for any compact semi-simple
gauge group.
The following general result was proved in \cite{Oliynyk1} for any compact semi-simple
gauge group and we have explicitly verified it for ${\mathfrak {su}}(N)$:
\begin{lemma}
\cite{Oliynyk1}
\begin{equation}
\frac {1}{2} {\bmath {\cal {W}}} = - \sum _{k=2}^{N} k\left( k - 1 \right) \bbeta _{k}(x) x^{k} +
\sum _{\ell =2}^{Z} \btau _{\ell }x^{\ell }
\end{equation}
for some $Z\in {\mathbb {N}}$ and vectors $\btau _{\ell }(x)$ which are regular at $x=0$.
\label{thm:Oliynyk1}
\end{lemma}
The key part of this result is that the expansion of ${\bmath {\cal {W}}}$ as a power
series about $x=0$ starts with terms of $O(x^{2})$, i.e. there are no $\ell =0$ or $\ell =1$
terms in this expansion.
The precise forms of the vectors $\btau _{\ell }$ are not particularly illuminating so we do not give them here.

Using Lemma \ref{thm:Oliynyk1} we have, for $k=2,\ldots, N$:
\begin{equation}
\frac {1}{2} {\bar {\bsigma }}_{k}^{T} {\bmath {\cal {W}}} =
-k \left( k - 1 \right) x^{k} \zeta _{k} +
\sum _{\ell =2}^{Z} {\bar {\bsigma }}_{k}^{T} \btau _{\ell }x^{\ell }.
\label{eq:btaubit1}
\end{equation}
To progress further, we need the following result on ${\bar {\bsigma }}_{k}^{T} \btau _{\ell }$, which is proved for
general gauge group in \cite{Oliynyk1}, and which we have verified directly for gauge group ${\mathfrak {su}}(N)$:
\begin{lemma}
\cite{Oliynyk1}
\begin{equation}
{\bar {\bsigma }}_{k}^{T} \btau _{\ell } = 0 \qquad
{\mbox {if $\, \ell < k +1 $.}}
\end{equation}
\label{thm:Oliynyk2}
\end{lemma}
Using Lemma \ref{thm:Oliynyk2}, equation (\ref{eq:btaubit1}) therefore simplifies to
\begin{equation}
\frac {1}{2} {\bar {\bsigma }}_{k}^{T} {\bmath {\cal {W}}} =
-k \left( k - 1 \right) x^{k} \zeta _{k+1} +
\sum _{\ell =k+1}^{Z} {\bar {\bsigma }}_{k}^{T} \btau _{\ell }x^{\ell }.
\end{equation}
The importance of this result will become apparent on returning to (\ref{eq:psi1}).
Defining a new variable $\chi _{k}(x)$ by $\chi _{k}(x) = \frac {d\zeta _{k}(x)}{dx}$, dividing throughout by
$x^{k+1}\mu $ and rearranging, we obtain
\begin{eqnarray}
x\frac {d\chi _{k}}{dx} & = &
-2 k \chi _{k} - k\left( k-1 \right) x^{-1} \zeta _{k} +
k \left( k-1 \right) x^{-1}\mu ^{-1} \zeta _{k}
\nonumber \\ & &
-\frac {1}{x^{2} \mu } \left[ 2{\tilde {m}} - \frac {2{\tilde {\Lambda }}x^{3}}{3}
- \frac {2x}{\lambda _{N} ^{2}}{\tilde {P}} \right]
\left[  k \zeta _{k} + x \chi _{k} \right]
\nonumber \\ & &
- \sum _{\ell = 0}^{Z-k-1} {\bar {\bsigma }}_{k}^{T} \btau _{\ell +k+1}x^{\ell } .
\label{eq:psi2}
\end{eqnarray}

Next we turn to the forms of the functions ${\tilde {G}}$ and ${\tilde {P}}$  (\ref{eq:tildeG&P}).
We write $\bbeta _{2}$ as
\begin{equation}
\bbeta _{2}(x) = \left[ {\hat {\beta }}_{0} + x{\hat {\beta }}_{1} (x) \right] {\bar {{\bmath {v}}}}_{2}
\label{eq:bbeta2}
\end{equation}
for some constant ${\hat {\beta }}_{0}$ and scalar function ${\hat {\beta }}_{1}(x)$,
where ${\bar {{\bmath {v}}}}_{2}$ is a right-eigenvector of the matrix $A$,
with eigenvalue $2$.
Once again, we make no assumption about the properties of ${\hat {\beta }}_{1}$.
We can write ${\hat {\beta }}_{1}$ as an analytic function of $\zeta _{2}$ using (\ref{eq:zetakdef}), but this
precise expression is not important here.
Using (\ref{eq:bbeta2}) and the properties of the right-eigenvector ${\bar {{\bmath {v}}}}_{2}$,
we are able to write ${\tilde {G}}$
and ${\tilde {P}}$ as follows:
\begin{equation}
{\tilde {G}} = \frac {2}{3} N \left( N^{2}-1 \right) x^{2} {\hat {\beta }}_{0}^{2} + x^{3} {\hat {G}};
\qquad
{\tilde {P}} = \frac {1}{3} N \left( N^{2} - 1\right) x^{2} {\hat {\beta }}_{0}^{2} + x^{3} {\hat {P}};
\end{equation}
where ${\hat {G}}$ is an analytic function of $x$, $\zeta _{k}$ and $\chi _{k}$,
and ${\hat {P}}$ is an analytic function of $x$ and the $\zeta _{k}$.
The differential equation satisfied by ${\tilde {m}}$ (\ref{eq:scaledequation1}) then simplifies to
\begin{equation}
\frac {d{\tilde {m}}}{dx} = 2x^{2}{\hat {\beta }}_{0}^{2} \left[ 2 \mu + 1 \right] + \frac {x^{3}}{\lambda _{N}^{2}}
\left[ \mu {\hat {G}} + {\hat {P}} \right] .
\end{equation}
We define yet another new variable $\alpha $ by
\begin{equation}
\alpha = \frac {1}{x^{3}} \left[ {\tilde {m}}-m_{3} x^{3} \right],
\end{equation}
where $m_{3} = 2{\hat {\beta }}_{0}^{2}$, and, after some algebra, we obtain the differential equation
satisfied by $\alpha $ in the form
\begin{equation}
x\frac {d\alpha }{dx} = -3\alpha + x{\cal {F}}_{\alpha },
\end{equation}
where ${\cal {F}}_{\alpha }$ is an analytic function in $\alpha $, $x$,
$\zeta _{k}$, $\chi _{k}$, and ${\tilde {\Lambda }}$.

Returning now to the YM equation (\ref{eq:psi2}), we next write
\begin{equation}
\mu ^{-1} = 1 + x^{2} {\tilde {\mu }}.
\end{equation}
Again, we make no assumptions about ${\tilde {\mu }}$: it can be written as a function of the
variables $x$ and $\alpha $, and the constants $m_{3}$ and ${\tilde {\Lambda }}$.
When we apply Theorem \ref{thm:BFM} at the end of this section, we will obtain as a corollary that
${\tilde {\mu }}$ is a regular function of $x$, at least in a neighbourhood of $x=0$.
The YM equation (\ref{eq:psi2}) then takes the form
\begin{equation}
x \frac {d\chi _{k}}{dx} = -2 k  \chi _{k} + x {\tilde {{\cal {F}}}}_{\chi _{k}}
+ \sum _{\ell = 0}^{Z-k-1} {\bar {\bsigma }}_{k}^{T} \btau _{\ell +k+1}x^{\ell } ,
\end{equation}
where ${\tilde {{\cal {F}}}}_{\chi _{k}}$ is an analytic function of the variables.
We make one last change of variables, defining \cite{Oliynyk1}
\begin{equation}
{\tilde {\chi }}_{k} = \chi _{k} + \frac {1}{2 k } {\bar {\bsigma }}_{k}^{T} \btau _{k+1}.
\end{equation}
The precise form of ${\bar {\bsigma }}_{k}^{T}\btau _{k+1}$ is not important, suffice to say it is a polynomial in
the $\zeta _{\ell }$ and does not involve any powers of $x$.

Then, summarizing, the field equations can be written as:
\begin{eqnarray}
x \frac {d\alpha }{dx} & = & -3\alpha + x{\cal {F}}_{\alpha };
\nonumber \\
x \frac {d\zeta _{k}}{dx} & = & x{\cal {F}}_{\zeta _{k}};
\nonumber \\
x \frac {d{\tilde {\chi }}_{k}}{dx} & = & -2 k  {\tilde {\chi }}_{k} +
x {\cal {F}}_{{\tilde {\chi }}_{k}};
\nonumber \\
x \frac {dS}{dx} & = & x{\cal {F}}_{S};
\nonumber \\
x \frac {d{\tilde {\Lambda }}}{dx} & = & 0;
\end{eqnarray}
where all the ${\cal {F}}$ functions are analytic in $x$, ${\tilde {\Lambda }}$, $\alpha $,
${\tilde {\chi }}_{k}$ and $\zeta _{k}$, in a neighbourhood of $\alpha = 0$, $x=0$.
Applying Theorem \ref{thm:BFM} we then have, in a neighbourhood of the origin $x=0$,
solutions of the form
\begin{eqnarray}
\alpha (x) & = & O(x); \qquad
\zeta _{k}(x) = \zeta _{0,k}+O(x); \qquad
{\tilde {\chi }}_{k} = O(x);
\nonumber \\
S(x) & = & S_{0} + O(x) ,
\end{eqnarray}
where $\zeta _{0,k}$ and $S_{0}$ are constants, and $\alpha $, $\zeta _{k}$, ${\tilde {\chi }}_{k}$ and
$S$ are analytic in $x$, ${\tilde {\Lambda }}$, $S_{0}$ and $\zeta _{0,k}$.
Returning to the original variables, we have proven the existence of solutions of the form
(\ref{eq:origin}) in a neighbourhood of the origin, which are analytic in $r$ and the parameters
${\bar {z}}_{k}$ and $S_{0}$.

\subsection{Local existence of solutions near the event horizon}
\label{sec:horizon}

As might be anticipated, the inclusion of the cosmological constant $\Lambda $ does not
change significantly the proof of the local existence of solutions near the event horizon $r=r_{h}$.
Our choice of variables follows those in \cite{Breitenlohner2,Winstanley2,Oliynyk1}.
For ${\mathfrak {su}}(2)$ EYM black holes in adS, the following result is stated in \cite{Winstanley1}.

\begin{proposition}
There exists an $N+2$-parameter family of local black hole solutions of the field equations
(\ref{eq:YMe},\ref{eq:Ee}) near $r=r_{h}>0$, satisfying the boundary conditions (\ref{eq:horizon}),
and analytic in $r_{h}$, $\Lambda $, $\omega _{j}(r_{h})$ and $r$.
\label{thm:horizon}
\end{proposition}

\paragraph{{\bf {Proof}}}

We define a new independent variable $x$ by $x=r-r_{h}$, and new dependent variables as follows
\cite{Breitenlohner2,Winstanley2,Oliynyk1}:
\begin{equation}
\rho = r, \qquad
\lambda = \frac {\mu }{x}, \qquad
\psi _{j} = \omega _{j}, \qquad
\xi _{j} = \frac {\mu \omega _{j}'}{x} = \lambda \omega _{j}'.
\label{eq:horvarsdef}
\end{equation}
The field equations (\ref{eq:YMe},\ref{eq:Ee}) then take the form
\begin{eqnarray}
x \frac {d\rho }{dx} & = & x;
\nonumber \\
x \frac {d\lambda }{dx} & = & - \lambda + x H_{\lambda } + F_{\lambda };
\nonumber \\
x \frac {d\psi _{j}}{dx} & = & \frac {x \xi _{j}}{\lambda };
\nonumber \\
x \frac {d\xi _{j}}{dx} & = & -\xi _{j} + x H_{\xi _{j}} + F_{\xi _{j}} ;
\nonumber \\
x \frac {dS}{dx} & = & \frac {2x}{\rho }GS;
\nonumber \\
x \frac {d\Lambda }{dx} & = & 0,
\label{eq:horexist1}
\end{eqnarray}
where
\begin{eqnarray}
F_{\lambda } & = &
\frac {1}{\rho } -\Lambda \rho - \frac {2}{\rho ^{3}}{\cal {P}} ;
\nonumber \\
H_{\lambda } & = &
-\frac {\lambda }{\rho }\left( 1 + 2G \right) ;
\nonumber \\
F_{\xi _{j}} & = &
-\frac {1}{\rho ^{2}} W_{j} \psi _{j};
\nonumber \\
H_{\xi _{j}} & = &
-\frac {2G}{\rho } \xi _{j};
\label{eq:horexist2}
\end{eqnarray}
we have defined a quantity ${\cal {P}}$ which is a polynomial in the $\psi _{j}$:
\begin{equation}
{\cal {P}} = r^{4} p_{\theta }
= \frac {1}{4}
\sum^N_{j=1}\left[\left(\psi^2_j-\psi^2_{j-1}-N-1+2j\right)^2\right]
\label{eq:calPdef}
\end{equation}
and $p_{\theta }$, $W_{j}$ and $G$ are given in (\ref{eq:ptheta},\ref{eq:Wdef},\ref{eq:Gdef}) respectively.
In particular, $G$ takes the form
\begin{equation}
G = \frac {1}{\lambda ^{2}} \sum _{j=1}^{N-1} \xi _{j}^{2}.
\end{equation}
The functions $F_{\lambda }$, $F_{\xi _{j}}$, $H_{\lambda }$ and $H_{\xi _{j}}$
are therefore polynomials in $1/\rho $, $1/\lambda $, $\rho $,
$\lambda $, $\psi _{j}$, $\xi _{j}$ and $\Lambda $.

The equations (\ref{eq:horexist1}) are not yet in the required form (\ref{eq:BFMsys}).
We require a further change of variables:
\begin{equation}
{\tilde {\lambda }} = \lambda - F_{\lambda }; \qquad
{\tilde {\xi }}_{j} = \xi _{j} - F_{\xi _{j}}.
\end{equation}
The differential equations for these new variables are:
\begin{eqnarray}
x \frac {d{\tilde {\lambda }}}{dx} & = & -{\tilde {\lambda }} + x G_{\lambda };
\nonumber \\
x \frac {d{\tilde {\xi _{j}}}}{dx} & = & -{\tilde {\xi }}_{j} + x G_{\xi _{j}},
\label{eq:horexist3}
\end{eqnarray}
where
\begin{equation}
G_{\lambda } = H_{\lambda } - \frac {dF_{\lambda }}{dx}; \qquad
G_{\xi _{j}} = H_{\xi _{j}} - \frac {dF_{\xi _{j}}}{dx}.
\end{equation}
We do not write out here the full forms of $G_{\lambda }$ and $G_{\xi _{j}}$, which are rather lengthy.
However, from the forms of $F_{\lambda }$ and $F_{\xi _{j}}$ (\ref{eq:horexist2}), using the differential
equations (\ref{eq:horexist1}), it is clear that both $G_{\lambda }$ and $G_{\xi _{j}}$ are
polynomials in $1/\rho $, $1/\lambda $, $\rho $, $\lambda $, ${\tilde {\lambda }}$, ${\tilde {\xi }}_{j}$
and $\Lambda $.

The equations (\ref{eq:horexist3}) are now in the form (\ref{eq:BFMsys})
required for the application of Theorem~\ref{thm:BFM}.
We have, in a neighbourhood of $x=0$, solutions of the form
\begin{eqnarray}
\rho & = & r_{h} + O(x); \qquad
{\tilde {\lambda }} = O(x); \qquad
\psi _{j} = \omega _{j}(r_{h}) + O(x);
\nonumber \\
{\tilde {\xi }}_{j} & = & O(x),
\qquad S = S(r_{h}) + O(x),
\end{eqnarray}
with $\rho $, ${\tilde {\lambda }}$, $\psi _{j}$ and ${\tilde {\xi }}_{j}$ all analytic in $x$, $r_{h}$,
$\omega _{j}(r_{h})$, $\Lambda $ and $S(r_{h})$.
Transforming back to our original variables, we have proven the existence of solutions of the field equations
in a neighbourhood of a black hole event horizon $r=r_{h}$, satisfying the boundary conditions (\ref{eq:horizon}),
and analytic in $r$, $r_{h}$, $\omega _{j}(r_{h})$, $\Lambda $ and $S(r_{h})$.

\subsection{Local existence of solutions near infinity}
\label{sec:infinity}

For asymptotically flat ${\mathfrak {su}}(N)$ solitons and black holes, the fact that the values of the
gauge field functions as $r\rightarrow \infty $ are fixed by (\ref{eq:AFinfinity}) makes the proof of local
existence of solutions in a neighbourhood of infinity rather complicated.
In fact, as shown in \cite{Oliynyk1,Kunzle2}, it is necessary to expand the gauge field functions $\omega _{j}(r)$
to order $r^{-N+1}$ in a manner similar to the analysis in Section \ref{sec:origin} near $r=0$.
In the asymptotically adS case, the fact that the values of the gauge field functions at infinity are not fixed
{\em {a priori}} makes the proof of the local existence of solutions much easier.
The result below is stated in the ${\mathfrak {su}}(2)$ case in \cite{Winstanley1}.

\pagebreak

\begin{proposition}
There exists an $2N$-parameter family of local solutions of the field equations
(\ref{eq:YMe},\ref{eq:Ee}) near $r=\infty $, satisfying the boundary conditions (\ref{eq:infinity}),
and analytic in $\Lambda $, $\omega _{j,\infty }$, $M$ and $r^{-1}$.
\end{proposition}

\paragraph{{\bf {Proof}}}

In this case our new independent variable is $x=r^{-1}$, and we define new dependent variables as follows,
following \cite{Breitenlohner2,Oliynyk1}:
\begin{equation}
\lambda = 2m = r\left( 1 - \mu - \frac {\Lambda r^{2}}{3} \right) ; \qquad
\psi _{j} =  \omega _{j} ; \qquad
\xi _{j} = r^{2} \omega _{j}' .
\end{equation}
Then the field equations (\ref{eq:YMe},\ref{eq:Ee}) take the form
\begin{eqnarray}
x \frac {d\lambda }{dx} & = & x f_{\lambda } ;
\nonumber \\
x \frac {d\psi _{j}}{dx} & = & -x \xi _{j};
\nonumber \\
x \frac {d\xi _{j}}{dx} & = & xf_{\xi _{j}};
\nonumber \\
x \frac {dS}{dx} & = & x^{4} f_{S};
\nonumber \\
x \frac {d\Lambda }{dx} & = & 0 .
\label{eq:infexist1}
\end{eqnarray}
The functions on the right-hand-side of (\ref{eq:infexist1}) are given by
\begin{eqnarray}
f_{\lambda } & = & -2 \left( x^{2} - x^{3} \lambda - \frac {\Lambda }{3} \right) \sum _{j=1}^{N-1} \xi _{j}^{2}
+ {\cal {P}} ;
\nonumber \\
f_{\xi _{j}} & = &
-\frac {3}{\Lambda } \left[ 1 - \frac {3x^{2}}{\Lambda } + \frac {3\lambda x^{3}}{\Lambda } \right] ^{-1}
\left[ W_{j} \omega _{j} - 2x \xi _{j} + 3\lambda x^{2} \xi _{j} - 2x^{3} \xi _{j} {\cal {P}} \right] ;
\nonumber \\
f_{S} & = & -2S \sum _{j=1}^{N-1} \xi _{j}^{2} ;
\end{eqnarray}
where ${\cal {P}} = x^{-4} p_{\theta }$ is a polynomial in the $\psi _{j}$, defined in equation (\ref{eq:calPdef}).
Immediately we have that the $f$'s are analytic in a neighbourhood of $x=0$, so applying Theorem~\ref{thm:BFM}
gives, in a neighbourhood of $x=0$, solutions of the form
\begin{eqnarray}
\lambda & = & 2M + O(x); \qquad
\psi _{j} = \omega _{j,\infty } + O(x); \qquad
\xi _{j} = c_{j} + O(x);
\nonumber \\
S & = & S_{\infty } + O(x^{4});
\end{eqnarray}
for constants $M$, $\omega _{j,\infty }$, $c_{j}$ and $S_{\infty }$,
and the solutions are analytic in $x$, $\Lambda $, and the constants.
If we fix $S_{\infty }=1$ for the space-time to be asymptotically anti-de Sitter, then the boundary conditions
(\ref{eq:infinity}) are satisfied,
and the field variables are analytic in $r^{-1}$, $\Lambda $, $M$, $\omega _{j,\infty }$ and $c_{j}$.
We note that we find additional free parameters in this case, corresponding to the coefficients of terms of $O(r^{-1})$
in the expansion of $\omega _{j}$ near infinity.
These parameters will not be important in our subsequent analysis.

\section{Existence of soliton and hairy black hole solutions }
\label{sec:existence}

In this section we prove two existence theorems: firstly, the existence of ${\mathfrak {su}}(N)$ solutions in
a neighbourhood of any embedded ${\mathfrak {su}}(2)$ solution; and secondly the existence of ${\mathfrak {su}}(N)$
solutions for which all the gauge field functions $\omega _{j}$ have no zeros provided $\left| \Lambda \right| $
is sufficiently large.

\subsection{Regularity for $\mu >0$}
\label{sec:simple}

Our strategy in proving the existence of genuinely
${\mathfrak {su}}(N)$ soliton and black hole solutions of the field equations is to
start with a local solution near the origin (in the soliton case) or black hole event horizon
and then extend this solution out to infinity.

The focus of this section is to show that solutions can be extended for larger values of $r$
provided that $\mu >0$.
This lemma was proved in \cite{Breitenlohner2} for ${\mathfrak {su}}(2)$ EYM with $\Lambda = 0$,
and stated for $\Lambda <0$ in \cite{Winstanley1}.
For ${\mathfrak {su}}(N)$ EYM with $\Lambda =0$, the corresponding lemma is stated in \cite{Winstanley2}.
Our method of proof is exactly the same as in \cite{Breitenlohner2}, so we only sketch the details.

\begin{lemma}
As long as $\mu >0$ all field variables are regular functions of $r$.
\label{thm:mu>0}
\end{lemma}

\paragraph{{\bf {Proof}}}

Consider an interval ${\cal {I}}=\left[ r_{1}, r_{2} \right) $, where $r_{1}<r_{2}$ and
$r_{1}>0$ in the soliton case, or $r_{1}>r_{h}$ in the black hole case.
Assume that all field variables are regular on the interval ${\cal {I}}$ and that $\mu (r)>0$
for all $r\in {\bar {\cal {I}}} = \left[ r_{1}, r_{2} \right] $.
We need to show that all other field variables (i.e. $\omega _{j}(r)$, $\omega _{j}'(r)$
and $S(r)$) remain regular at $r_{2}$.

From the fact that $\mu (r_{2})>0$ we obtain
\begin{equation}
r_{2} - \frac {\Lambda r_{2}^{3} }{3} > 2m(r_{2}).
\end{equation}
From (\ref{eq:Ee}), since $r^{2} p_{\theta }$ is positive,
we have
\begin{equation}
2m(r_{2}) \ge 2 \int _{r_{1}}^{r_{2}} \mu (r) \sum _{j=1}^{N-1} \omega _{j}'^{2} \  dr.
\end{equation}
Now $\mu (r)$ must have a minimum in the closed interval ${\bar {\cal {I}}}$, so defining
\begin{equation}
\mu _{\rm {min}} = \min \left\{ \mu (r): r\in {\bar {\cal {I}}} \right\} >0;
\end{equation}
we find
\begin{equation}
\int _{r_{1}}^{r_{2}} \sum _{j=1}^{N-1} \omega _{j}'^{2} \, dr  < \frac {1}{2\mu _{\rm {min}}}
\left( r_{2} - \frac {\Lambda r_{2}^{3}}{3} \right) .
\end{equation}
Direct integration of the second Einstein equation (\ref{eq:Ee}) then gives that $\log S$ and therefore $S$
is finite at $r_{2}$.

Using the Cauchy-Schwarz inequality, we have
\begin{equation}
\sum _{j=1}^{N-1} \left[ \omega _{j}(r_{1}) - \omega _{j}(r_{2}) \right] ^{2}
 = \sum _{j=1}^{N-1} \left[ \int _{r_{1}}^{r_{2}} \omega _{j}' \, dr \right] ^{2}
 \le \left( r_{2} - r_{1} \right)  \int _{r_{1}}^{r_{2}} \sum _{j=1}^{N-1} \omega _{j}'^{2} \, dr .
\end{equation}
Since the left-hand-side of the above equation is a sum of positive terms, and the right-hand-side is bounded above,
we deduce that each $\omega _{j}(r_{2})$ is finite.

Finally, we need to show that each $\omega _{j}'(r_{2})$ is finite.
To do this, we write the Yang-Mills equation (\ref{eq:YMe}) in the form
\begin{equation}
\left( \mu \omega _{j}' \right) ' + \frac {S'}{S} \mu \omega _{j}' = - \frac {W_{j}\omega _{j}}{r^{2}}.
\end{equation}
Integrating this first order differential equation for $\mu \omega _{j}'$ gives
\begin{equation}
\omega _{j}'(r_{2}) =
\frac {1}{S(r_{2})\mu (r_{2})} \left[
S(r_{1})\mu (r_{1}) \omega _{j}'(r_{1}) - \int _{r_{1}}^{r_{2}} \frac {SW_{j}\omega _{j}}{r^{2}} \, dr \right] ,
\end{equation}
so that $\omega _{j}'(r_{2})$ is finite and we have shown that all field variables are finite on the closed
interval ${\bar {\cal {I}}}$.

\subsection{Asymptotic behaviour as $r\rightarrow \infty $}
\label{sec:asympt}

One of the most important reasons for the abundance of solutions of ${\mathfrak {su}}(2)$ EYM with
$\Lambda <0$ compared with the asymptotically flat case is the difference in the behaviour of the Yang-Mills
equations as $r\rightarrow \infty $.
Therefore we next turn to the asymptotic behaviour of the ${\mathfrak {su}}(N)$ Yang-Mills equations (\ref{eq:YMe}).

As $r\rightarrow \infty $, these take the form
\begin{equation}
-\frac {\Lambda r^{4}}{3} \omega _{j} '' - \frac {2\Lambda r^{3}}{3} \omega _{j}' + W_{j}\omega _{j}=0.
\label{eq:YMinfinity}
\end{equation}
As in the ${\mathfrak {su}}(2)$ case \cite{Winstanley1}, these can be made autonomous by the change of variable
\begin{equation}
\tau = \frac {1}{r} {\sqrt {-\frac {\Lambda }{3}}} ,
\label{eq:taudef}
\end{equation}
the Yang-Mills equations becoming
\begin{equation}
\frac {d^{2} \omega _{j}}{d\tau ^{2}} + W_{j} \omega _{j} =0.
\label{eq:phaseAdS}
\end{equation}
The autonomous equations (\ref{eq:phaseAdS}) have critical points when
\begin{equation}
W_{j} \omega _{j} = \left( 1- \omega _{j}^{2} + \frac {1}{2} \omega _{j+1}^{2} + \frac {1}{2} \omega _{j-1}^{2}
\right) \omega _{j} = 0.
\end{equation}
If all the $\omega _{j}$ are non-zero at a critical point, their values are:
\begin{equation}
\omega _{j} = \pm {\sqrt {j \left( N - j \right) }} .
\end{equation}
Linearization about the critical points reveals that the critical point at the origin is a centre, and all
other critical points are saddles.

We are in exactly the same situation as in the ${\mathfrak {su}}(2)$ case: what is important is not so much
the different critical point structure of the equation (\ref{eq:phaseAdS}) compared with the asymptotically flat
case \cite{Winstanley2}, but rather the nature of the variable $\tau $ (\ref{eq:taudef}).
In particular, when $\Lambda <0$ the variable $\tau $ tends to zero as $r \rightarrow \infty $, so that if we
consider the solutions of the equations (\ref{eq:phaseAdS}) starting at some large value of $r=r_{1}$
and going all the way out to infinity, the corresponding values of $\tau $ will lie in some small interval
$\tau \in \left[ 0, \tau _{1} \right] $.
On the other hand, for asymptotically flat solutions, the variable ${\tilde {\tau }}$ which makes
the Yang-Mills equations in flat space autonomous is ${\tilde {\tau }}=\log r$, which will tend to infinity
as $r\rightarrow \infty $.
Therefore, in the asymptotically flat case, any solution of the Yang-Mills equations must trace out the whole length
of a phase path, whereas in the asymptotically adS case the solutions will only travel a small, finite
distance along any one phase path.

Including a small gravitational perturbation into the Yang-Mills equations (\ref{eq:YMinfinity})
will not change this key property of the variable $\tau $ nor the finiteness of the phase paths taken by the
field variables as $r\rightarrow \infty $.

\subsection{Existence of ${\mathfrak {su}}(N)$ solutions in a neighbourhood of embedded ${\mathfrak {su}}(2)$
solutions}
\label{sec:result}

We now have available all the ingredients we require to show the existence of genuinely ${\mathfrak {su}}(N)$
EYM solitons and black holes.
In this subsection we prove our first existence theorem, namely the existence of ${\mathfrak {su}}(N)$
solutions in a neighbourhood of an embedded ${\mathfrak {su}}(2)$ solution.
This result relies heavily on the analyticity of the local solutions, proved in Section \ref{sec:local}.

\begin{proposition}
Suppose there is an embedded ${\mathfrak {su}}(2)$ solution of the field equations, with
the gauge field functions all having $k$ zeros.
Then, all initial
parameters in a sufficiently small neighbourhood of the initial parameters giving the
embedded ${\mathfrak {su}}(2)$ solution will give an ${\mathfrak {su}}(N)$ solution
of the field equations in which all gauge field functions have $k$ zeros.
\end{proposition}

\paragraph{{\bf {Proof}}}

Suppose we have an embedded, non-trivial, ${\mathfrak {su}}(2)$ solution of the field equations.
If this is a soliton, the initial conditions at the origin will be given by (\ref{eq:origin}) with
${\bar {z}}_{2}\neq 0$ and ${\bar {z}}_{3}={\bar {z}}_{4} = \ldots = {\bar {z}}_{N}=0$.
If this is a black hole, the initial conditions at the event horizon will be $\omega _{1}(r_{h}) =
\omega _{2}(r_{h}) = \ldots = \omega _{N-1}(r_{h}) \neq 0$.
From these initial conditions, we can integrate the field equations (\ref{eq:YMe},\ref{eq:Ee}) all the
way out to infinity to give a regular solution.
Furthermore, the solution will satisfy the boundary conditions (\ref{eq:infinity}) at infinity.
For the rest of this section, we will assume that the magnitude of the cosmological constant $\Lambda $ is fixed,
and, if we are considering black hole solutions, that the radius of the event horizon, $r_{h}$, is also fixed.
We also suppose that the gauge field functions each have $k$ zeros (they will all have the same number of zeros
from (\ref{eq:embeddedsu2}), since this is an embedded ${\mathfrak {su}}(2)$ solution).

From the local existence theorems (Propositions \ref{thm:origin} and \ref{thm:horizon} proved
in Sections \ref{sec:origin} and \ref{sec:horizon} respectively) we know that there are solutions of the field
equations locally near the origin or event horizon as applicable, for any values of the initial parameters
${\bar {z}}_{2}, \ldots {\bar {z}}_{N}$ or $\omega _{1}(r_{h}),\ldots \omega _{N-1}(r_{h})$, and that these solutions
are analytic in these initial parameters.
For the embedded ${\mathfrak {su}}(2)$ solution, it must be the case that $\mu (r)>0$ for all
$r\in \left[ r_{0},\infty \right) $ (where $r_{0}=0$ for solitons and $r_{0}=r_{h}$ for black holes).
Therefore, by analyticity, for initial parameters close to the initial parameters giving the embedded
${\mathfrak {su}}(2)$ solution, the corresponding local solutions will have $\mu (r)>0$ for all
$r \in \left[ r_{0},r_{2} \right] $ for some $r_{2}$.
By Lemma \ref{thm:mu>0}, these local solutions will also be regular on the interval $\left[ r_{0}, r_{2} \right] $.

Now fix some $r_{1} \gg \max \{ 1,r_{0} \}$, such that $\left| \Lambda \right| r_{1}^{2} \gg 1$ and such that,
for the embedded ${\mathfrak {su}}(2)$ solution, we have $m(r_{1})/r_{1} \ll 1$.
By the analyticity argument above, provided the initial parameters
${\bar {z}}_{2}, \ldots ,{\bar {z}}_{N}$ or $\omega _{1}(r_{h}),\ldots \omega _{N-1}(r_{h})$ (as applicable)
are in a sufficiently small neighbourhood of the initial parameters giving the embedded ${\mathfrak {su}}(2)$
solution, the local ${\mathfrak {su}}(N)$ solutions
whose existence is guaranteed by Propositions \ref{thm:origin} and \ref{thm:horizon}
will be regular on the whole interval $r\in \left[ r_{0}, r_{1} \right] $.
That is, in this interval $\mu >0$ for all $r$ and the gauge field functions $\omega _{j}$ will all have
$k$ zeros provided $r_{1}$ is larger than the largest zero of the gauge field function $\omega $ of the
embedded ${\mathfrak {su}}(2)$ solution.
Furthermore, at $r_{1}$, it will still be the case that $m(r_{1})/r_{1} \ll 1$ for these ${\mathfrak {su}}(N)$
solutions as well as the nearby embedded ${\mathfrak {su}}(2)$ solution.

Since $r_{1} \gg \max \{ 1, r_{0} \}$ and $m(r_{1})/r_{1} \ll 1$ for these ${\mathfrak {su}}(N)$ solutions,
we can use the asymptotically adS regime discussed in Section \ref{sec:asympt}.
Provided $r_{1}$ is sufficiently large, the solutions will not move very far along their phase path as $r$ increases
from $r_{1}$ to $r\rightarrow \infty $.
In particular, the gauge field functions will have no further zeros, and both the gauge field
functions and their derivatives will not vary very much from their values at $r=r_{1}$.
This means that $m(r)/r$ will continue to be very small as the solutions move along their phase path, and the
asymptotic adS regime will continue to be valid.

\bigskip

Therefore, we have shown the existence of genuinely ${\mathfrak {su}}(N)$ soliton and black hole solutions
in a neighbourhood of any embedded ${\mathfrak {su}}(2)$ solution.
The gauge field functions of these ${\mathfrak {su}}(N)$ solutions will have the same number of zeros as the
gauge field functions of the embedded ${\mathfrak {su}}(2)$ solution.
It should be emphasized that these are genuinely ${\mathfrak {su}}(N)$ solutions rather than embedded
${\mathfrak {su}}(2)$ solutions because the gauge field functions $\omega _{j}(r)$ will not satisfy
(\ref{eq:embeddedsu2}).

This result can be applied to the embedded Schwarzschild-adS or pure adS solutions (with initial parameters,
$\omega _{j}(r_{h}) = {\sqrt {j\left( N-j \right) }}$ and ${\bar {z}}_{j}=0$ respectively) to give, for {\em {any}}
negative value of the cosmological constant $\Lambda $, genuinely ${\mathfrak {su}}(N)$ solutions for which all
the gauge field functions have no zeros.

\begin{corollary}
For any $\Lambda <0$, there exist ${\mathfrak {su}}(N)$ black holes and solitons for which all the
gauge field functions have no zeros.
\label{thm:corollary}
\end{corollary}

In practice, the size of this neighbourhood about the embedded pure adS or Schwarzschild-adS solutions
in which we have ${\mathfrak {su}}(N)$ soliton and black hole solutions where the gauge
field functions have no zeros is negligibly small unless
\begin{equation}
\left| \Lambda \right| \gtrsim 10 ^{-1} \lambda _{N}^{2} =  \frac {10^{-1}}{6} N (N-1)(N+1) .
\end{equation}
For the ${\mathfrak {su}}(2)$ case, this can be seen in Figures 1 and 4 in \cite{Baxter2}.
The factor of $\lambda _{N}^{2}$ arises from the embedding of the ${\mathfrak {su}}(2)$ solutions in
${\mathfrak {su}}(N)$ (see Section \ref{sec:su2embedded}), and means that the size of the
neighbourhood decreases as $N$ increases.
This is illustrated in Figures 9 and 13 in \cite{Baxter2}, where the size of the region
of parameter space where we have nodeless black hole solutions with $r_{h}=1$ and $\Lambda = -1$ can
be compared for ${\mathfrak {su}}(3)$ and ${\mathfrak {su}}(4)$ EYM, the latter region being significantly
smaller than the former.

\subsection{Existence of solutions for sufficiently large $\left| \Lambda \right| $}
\label{sec:largeL}

In the previous section we have proven the existence of black hole and soliton solutions of the
${\mathfrak {su}}(N)$ EYM field equations, for any negative value of the cosmological constant $\Lambda $,
such that all the gauge field functions have no zeros.
However, the size of the region of parameter space where we have these solutions shrinks as $N$ increases,
for fixed $\left| \Lambda \right| $.
In this section we shall consider the behaviour of the solutions for large $\left| \Lambda \right| $.
We shall prove that, given fixed initial parameters at the origin or black hole event horizon, as
applicable,
for sufficiently large $\left| \Lambda \right| $, the corresponding
solutions of the field equations are such that all the gauge field functions have no zeros.
These solutions are of particular interest because we expect \cite{Baxter1} that at least some
of them will be linearly stable.

Numerically (see Section \ref{sec:numerics} and \cite{Baxter2}), we find that for $\left| \Lambda \right| $
sufficiently large, {\em {all}} soliton and black hole solutions of the field equations are such
that all the gauge field functions have no zeros.
This is not what we are able to prove in this section.
Here we can show that, with fixed initial parameters, there are nodeless solutions for sufficiently large
$\left| \Lambda \right| $, and it may well be that the magnitude of $\left| \Lambda \right| $ required
for the existence of nodeless solutions varies depending on the values of the initial parameters.
We observe this numerically, for example, comparing Figures \ref{fig:su3bhlambda-2} and \ref{fig:su3bhlambda-3}
in the ${\mathfrak {su}}(3)$ case,
we see that for $\omega _{1}(r_{h})=0.25$, $\omega _{2}(r_{h})=1$, the corresponding black hole solution is not
nodeless when $\Lambda = -2$ but is nodeless when $\Lambda = -3$, whereas for $\omega _{1}(r_{h})=0.75$,
$\omega _{2}(r_{h})=1$, the black hole is nodeless for both these values of $\Lambda $.

We need to consider black hole and soliton solutions separately.

\subsubsection{Black holes}
\label{sec:largeLbh}

In \cite{Winstanley1}, for ${\mathfrak {su}}(2)$ EYM black holes, it is shown that for any fixed value of the
event horizon radius $r_{h}$ and any fixed value of the
gauge field function on the horizon, $\omega (r_{h})$, then for all $\Lambda $ such that $\left| \Lambda \right| $ is
sufficiently large, there is a black hole solution of the field equations such that the gauge field function
$\omega (r)$ has no zeros.

The extension of this result to ${\mathfrak {su}}(N)$ EYM black holes, as might be expected, is not difficult so
we just briefly sketch the derivation of the following result.

\begin{proposition}
For fixed $r_{h}$ and any fixed values of the gauge field functions at the event horizon, $\omega _{j}(r_{h})$,
for $\left| \Lambda \right| $ sufficiently large, there exists a black hole solution of the ${\mathfrak {su}}(N)$
EYM field equations such that all the gauge field functions $\omega _{j}(r)$ have no zeros.
\label{thm:bhlargeL}
\end{proposition}

\paragraph{{\bf {Proof}}}

Firstly, we note that for fixed $r_{h}$ and $\omega _{j}(r_{h})$, the constraint (\ref{eq:constraint})
for a regular event horizon is satisfied for all sufficiently large $\left| \Lambda \right| $.
As in \cite{Winstanley1}, it is helpful to define
a length scale $\ell $ by $\ell ^{2} = -3/\Lambda $, and then
new variables ${\hat {m}}$ and ${\hat {\mu }}$,
which will be finite as $\left| \Lambda \right| \rightarrow \infty $, $\ell \rightarrow 0$
as follows:
\begin{equation}
{\hat {m}} = m \ell ^{2}, \qquad
{\hat {\mu }} = \mu  \ell ^{2} = \ell ^{2} - \frac {2{\hat {m}}}{r} + r^{2}.
\end{equation}
The field equations (\ref{eq:YMe},\ref{eq:Ee}) then take the form
\begin{eqnarray}
{\hat {m}}' & = & \left( \ell ^{2} - \frac {2{\hat {m}}}{r} + r^{2} \right) G
+ \ell ^{2} r^{2} p_{\theta };
\qquad
\frac {S'}{S} = \frac {2G}{r};
\nonumber \\
0 & = &
r^{2} \left( \ell ^{2} - \frac {2{\hat {m}}}{r} + r^{2} \right) \omega _{j}''
+ \left[ 2{\hat {m}} - 2\ell ^{2}r^{3}p_{\theta } + 2r^{3} \right] \omega _{j}'
+ \ell ^{2} W_{j} \omega _{j}.
\end{eqnarray}
In the limit $\ell \rightarrow 0$, these equations simplify considerably and have the unique solution
\begin{equation}
{\hat {m}}(r) = {\hat {m}}(r_{h}) = \frac {1}{2} r_{h}^{3}; \qquad
S (r) = 0; \qquad
\omega _{j} (r) = \omega _{j}(r_{h}).
\end{equation}

We would like to extend these results to small, non-zero values of $\ell $ by analyticity,
along the lines of the argument used in Section \ref{sec:result}.
The proof of Proposition \ref{thm:horizon}, specifically contains $\Lambda $, however,
we may use the straightforward change of variables (cf. (\ref{eq:horvarsdef})):
\begin{equation}
{\hat {\lambda }} = \ell ^{2} {\lambda }.
\end{equation}
Then the equations (\ref{eq:horexist1}) are unchanged except that we include $x\frac {d\ell }{dx}=0$ instead
of $x\frac {d\Lambda }{dx}=0$ and have the following equation for $x\frac {d{\hat {\lambda }}}{dx}$:
\begin{equation}
x \frac {d{\hat {\lambda }}}{dx} = -{\hat {\lambda }} + xH_{\hat {\lambda }} + F_{\hat {\lambda }};
\end{equation}
where
\begin{eqnarray}
H_{\hat {\lambda }} & = &
-\frac {{\hat {\lambda }}}{\rho } \left( 1+ 2G \right) ;
\nonumber \\
F_{\hat {\lambda }} & = & \frac {\ell ^{2}}{\rho } + 3\rho - \frac {2\ell ^{2}}{\rho ^{3}} {\cal {P}}.
\end{eqnarray}
The field equations are then all regular as $\ell \rightarrow 0 $, and the local existence result
Proposition \ref{thm:horizon} carries over to give regular solutions in a neighbourhood of the event horizon.
Furthermore, these solutions will be analytic in $r_{h}$, $\omega _{j}(r_{h})$ and $\ell $.

Once we have the local existence of solutions near the event horizon, exactly the same argument
as used in Section \ref{sec:result} shows that we have solutions for $\ell $ sufficiently small.
In particular, fixing some $r_{1}\gg r_{h}$, and fixing both the radius of the event horizon and
the values of the gauge field functions on the event horizon, and varying just $\ell $,
for $\ell $ sufficiently small we will have local solutions near the event horizon which are regular for
all $r\in \left[ r_{h}, r_{1} \right] $ and for which all the gauge field functions have no zeros in this interval.
Then, provided that $r_{1}$ is sufficiently large, we may use the asymptotic regime (see Section \ref{sec:asympt}),
and find that, as $r\rightarrow \infty $, these solutions remain regular and the gauge field functions will
have no zeros.

\begin{figure}[p]
\begin{center}
\includegraphics[width=8cm,angle=270]{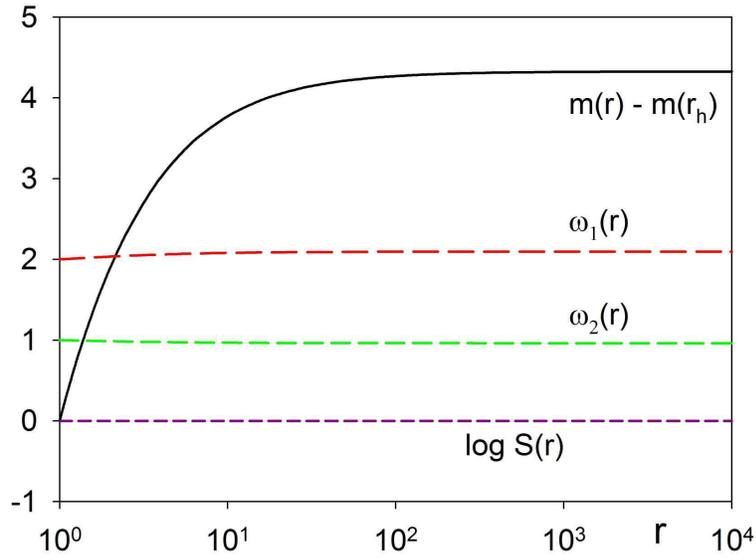}
\end{center}
\caption{Example of an ${\mathfrak {su}}(3)$ black hole solution with $r_{h}=1$, $\Lambda = -100$,
$\omega _{1}(r_{h}) = 2$ and $\omega _{2}(r_{h})=1$.
As $r\rightarrow \infty $, the gauge field functions tend to the following limits:
$\omega _{1}\rightarrow 2.0962$, $\omega _{2} \rightarrow 0.9625$.
The function $\log S$ is not identically zero: at the horizon $\log S = 2.6246 \times 10^{-3}$.
}
\label{fig:su3bhlargeLexample1}
\end{figure}
\begin{figure}[p]
\begin{center}
\includegraphics[width=8cm,angle=270]{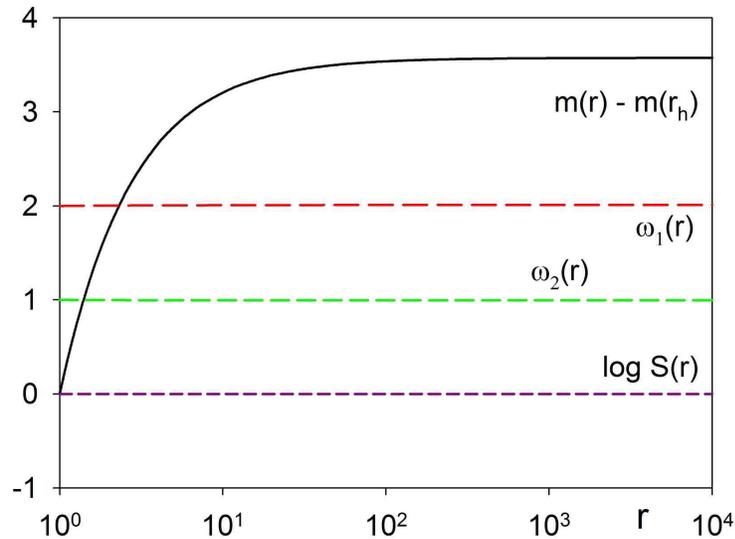}
\end{center}
\caption{Example of an ${\mathfrak {su}}(3)$ black hole solution with $r_{h}=1$, $\Lambda = -1000$,
$\omega _{1}(r_{h}) = 2$ and $\omega _{2}(r_{h})=1$.
As $r\rightarrow \infty $, the gauge field functions tend to the following limits:
$\omega _{1}\rightarrow 2.0091$, $\omega _{2} \rightarrow 0.9964$.
The function $\log S$ is not identically zero: at the horizon $\log S = 2.3968 \times 10^{-5}$.
}
\label{fig:su3bhlargeLexample2}
\end{figure}

\bigskip

The results of this section are illustrated in Figures \ref{fig:su3bhlargeLexample1} and
\ref{fig:su3bhlargeLexample2}, where we have plotted example black hole solutions with the same values of
the event horizon radius $r_{h}$ and values of the gauge field functions at the event horizon, $\omega _{1}(r_{h})$
and $\omega _{2}(r_{h})$, but with $\Lambda = -100$ and $\Lambda = -1000$ respectively.

The effect of increasing $\left| \Lambda \right| $ by an order of magnitude can be seen by comparing
these two figures.
In both cases, the gauge field functions hardly vary at all from their values at the event horizon, and the
difference between their values at the event horizon and at infinity reduces as $\left| \Lambda \right| $
increases.
In both figures $\log S$ is a small function, and it decreases as $\left| \Lambda \right| $ increases.
One interesting point is that, although we are considering very large values of $\left| \Lambda \right| $ in both
these figures, the geometry is far from being Schwarzschild-adS.
This is shown by the fact that the metric function $m(r)$ is far from being constant, although the difference between
its values on the event horizon and at infinity is decreasing as $\left| \Lambda \right| $ increases, albeit slowly.
Of course, Proposition \ref{thm:bhlargeL} tells us that $m(r)$ will be approximately constant only for ``sufficiently
large'' $\left| \Lambda \right| $, and does not tell us how large ``sufficiently large'' is.
It is clear that $\left| \Lambda \right| $ will have to be very large indeed for $m(r)$ to be approximately constant,
although the gauge field functions become approximately constant for much smaller values of $\left| \Lambda \right| $.
However, our main interest in this paper is proving the existence of solutions, even if in an extreme
situation.

\subsubsection{Solitons}
\label{sec:largeLsol}

Proving the existence of soliton solutions for {\em {any}} values of the initial parameters
$\left( {\bar {z}}_{2}, \ldots , {\bar {z}}_{N} \right)$ (\ref{eq:origin}),
for sufficiently large $\left| \Lambda \right| $,
is much more difficult than the corresponding result in the black hole case.
The reason for this is that, in the black hole case, there are two length scales: the radius of the event horizon $r_{h}$
and the length scale $\ell $ set by the negative cosmological constant ($\ell ^{2} = -3/\Lambda $),
and we have $r\ge r_{h}$ for the region of space-time outside the event horizon.
Therefore, with $r_{h}$ fixed, we can take $\ell \rightarrow 0$ in a reasonably straightforward way because we
know that $r^{-1}$ is bounded.
However, for soliton solutions, there is only one length scale, namely $\ell $, and we need to consider the whole range
of values of $r\in \left[ 0, \infty \right) $.
We therefore have to be very much more careful in how we take the limit $\ell \rightarrow 0$.

For solitons, the result corresponding to Proposition \ref{thm:bhlargeL} is:

\begin{proposition}
For any fixed values of parameters ${\bar {z}}_{i}$, $i=2,\ldots , N$, which parameterize the gauge field functions
near the origin,
for $\left| \Lambda \right| $ sufficiently large, there exists a soliton solution of the ${\mathfrak {su}}(N)$
EYM field equations such that all the gauge field functions $\omega _{j}(r)$ have no zeros.
\label{thm:solitonlargeL}
\end{proposition}

\paragraph{{\bf {Proof}}}

Firstly, we rescale all the dimensionful quantities as follows:
\begin{equation}
r=\ell {\hat {x}}; \qquad m(r) = \ell {\hat {m}}({\hat {x}}).
\label{eq:hatx}
\end{equation}
In the limit $\ell \rightarrow 0$, it is more convenient to write the Yang-Mills equations not in terms
of the gauge field functions $\omega _{j}$ (\ref{eq:YMe}), but instead to write the gauge field functions
in terms of new functions ${\hat {\zeta }}_{j}(r)$:
\begin{equation}
\bomega (r) = \bomega _{0} + \sum _{k=2}^{N} {\hat {\zeta }}_{k} {\bmath {v}}_{k} r^{k}
= \bomega _{0} + \sum _{k=2}^{N} {\hat {\zeta }}_{k}({\hat {x}}) \ell ^{k} {\hat {x}}^{k} {\bmath {v}}_{k}.
\label{eq:omegainfinity}
\end{equation}
These functions ${\hat {\zeta }}_{k}$ are essentially the same as the $\zeta _{k}$ in Section \ref{sec:originproof},
modulo a factor of ${\sqrt {j(N-j)}}$ and
the rescaling of the independent variable $x$ (compare equations (\ref{eq:xorigin}) and (\ref{eq:hatx})).
As shown in Section \ref{sec:originproof}, the differential equations satisfied by the ${\hat {\zeta }}_{k}(x)$
(\ref{eq:psi1}) are rather complicated for general $N$.
For $N=3$, they are given explicitly in \cite{Baxter2}.
However, in the limit $\ell \rightarrow 0$, the field equations simplify considerably.
The Einstein equations reduce to
\begin{equation}
{\hat {m}} \equiv 0, \qquad S \equiv 1,
\end{equation}
so that $\mu = 1+ x^{2}$,
and the differential equations for the ${\hat {\zeta }}_{k}(x)$ (\ref{eq:psi1}) decouple:
\begin{eqnarray}
0 & = & x\left( 1 + x^{2} \right) \frac {d^{2}{\hat {\zeta }}_{k}}{dx^{2}} + 2\left[
 k  + \left( k+1 \right) x^{2} \right] \frac {d{\hat {\zeta }}_{k}}{dx}
\nonumber \\ & &
+ x k  \left( k+1 \right) {\hat {\zeta }}_{k}.
\label{eq:zetainfinity}
\end{eqnarray}
This equation has a solution which is regular at $x=0$:
\begin{equation}
{\hat {\zeta }}_{k}(x) \propto {}_{2}F_{1} \left( \frac {1}{2} + \frac {k}{2} , \frac {k}{2} ;
\frac {1}{2} + k ; -x^{2} \right) ;
\label{eq:zetahpg}
\end{equation}
where ${}_{2}F_{1}$ is a hypergeometric function.
The constant of proportionality in (\ref{eq:zetahpg}) is simply ${\bar {z}}_{k}$ (\ref{eq:origin}).
It is straightforward to show, using the properties of hypergeometric functions, that the boundary
conditions at infinity (\ref{eq:infinity}) are satisfied.
From (\ref{eq:omegainfinity}), when $\ell =0$, all the gauge field functions are constant and given by
$\omega _{j}\equiv \pm {\sqrt {j(N-j)}}$.
However, it is important to find the forms of the ${\hat {\zeta }}_{k}$ functions in the limit $\ell \rightarrow 0$,
even though they are multiplied by powers of $\ell $ and so do not affect the form of the gauge field functions
in this limit.
This is because they are important for small, non-zero $\ell $.

The local existence result (Proposition \ref{thm:origin}) of Section \ref{sec:originproof}
carries over, with trivial amendments, using the
scaled radial co-ordinate ${\hat {x}}$ rather than $x$.
Therefore in a neighbourhood of the origin, we have local solutions of the field equations which are analytic
in the parameters ${\bar {z}}_{k}$ and $\ell $.
The same argument as used in Sections \ref{sec:result} and \ref{sec:largeLbh} can then easily be used
to show that there are soliton solutions for the field equations for sufficiently small $\ell $ (that is,
sufficiently large $\left| \Lambda \right| $) for which all the gauge field functions have no zeros.

\bigskip

\begin{figure}
\begin{center}
\includegraphics[width=8cm,angle=270]{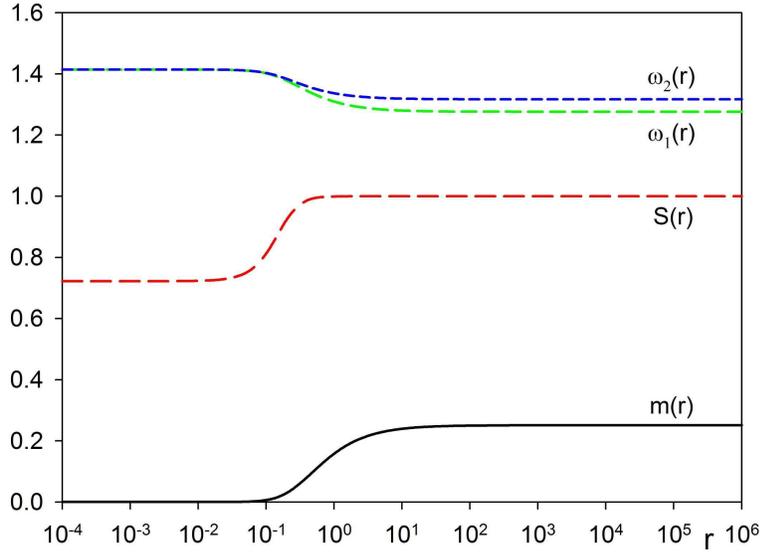}
\caption{Example of an ${\mathfrak {su}}(3)$ soliton solution with $\Lambda = -100$,
${\bar {z}}_{2} = -2$ and ${\bar {z}}_{3}=-1$. }
\label{fig:su3solitonlargeLexample1}
\end{center}
\end{figure}

\begin{figure}
\begin{center}
\includegraphics[width=8cm,angle=270]{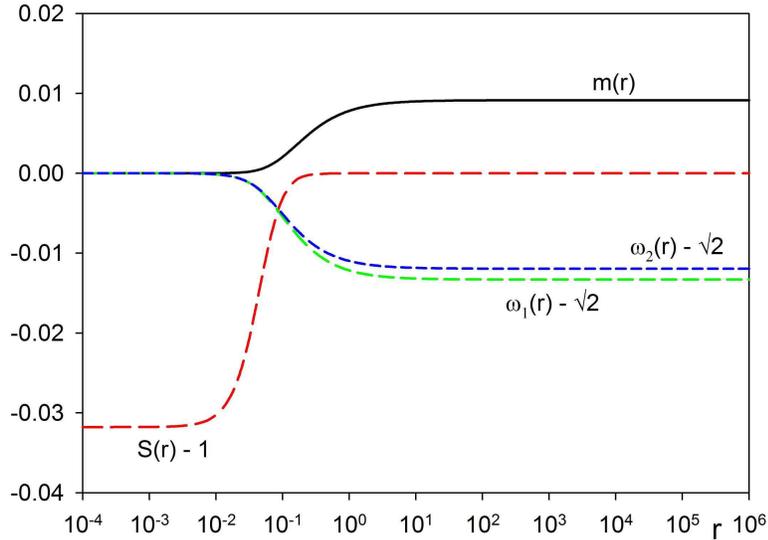}
\caption{Example of an ${\mathfrak {su}}(3)$ soliton solution with $\Lambda = -1000$,
${\bar {z}}_{2} = -2$ and ${\bar {z}}_{3}=-1$.}
\label{fig:su3solitonlargeLexample2}
\end{center}
\end{figure}

As with the black hole solutions in the previous section, we illustrate this result with two example
soliton solutions in Figures \ref{fig:su3solitonlargeLexample1} and \ref{fig:su3solitonlargeLexample2}.
Comparing these two figures, it can be seen how $m(r)$ and $\log S(r)$ tend to zero as $\left| \Lambda \right| $ increases,
and the gauge field functions $\omega _{1,2}$ approach ${\sqrt {2}}$ everywhere.
The metric function $m(r)$ is decreasing in size more quickly for solitons as $\left| \Lambda \right| $ increases,
compared with the black hole case (cf. Figures \ref{fig:su3bhlargeLexample1} and \ref{fig:su3bhlargeLexample2}).
Our results in this section agree with \cite{Hosotani}, where it is shown that stable monopoles
(in \cite{Hosotani} both monopoles and dyons are considered, but here we have studied only purely magnetic
monopole configurations) in ${\mathfrak {su}}(2)$ EYM with $\Lambda <0$ are approximated by solutions
on a pure adS background.
There are non-trivial, nodeless, monopole solutions in pure adS (see Figure 1 in \cite{Hosotani}) for
the ${\mathfrak {su}}(2)$ case, and, as in Section \ref{sec:su2embedded}, these can be embedded into
${\mathfrak {su}}(N)$.
In Proposition \ref{thm:solitonlargeL}, we have shown the existence of nodeless ${\mathfrak {su}}(N)$
solitons effectively approximated by solitons in pure adS (because ${\hat {m}} \approx 0$),
but only in the limit of very large $\left| \Lambda \right| $, in which case the equations
governing the YM degrees of freedom decouple (\ref{eq:zetainfinity}).


\section{Conclusions}
\label{sec:conc}

In this paper we have studied the existence of four-dimensional, spherically symmetric,
soliton and black hole solutions of ${\mathfrak {su}}(N)$
EYM with a negative cosmological constant.
Numerical solutions for $N=3$ and $N=4$ were presented in \cite{Baxter2}, but, of course,
numerics can only find solutions for a small, finite number of values of $N$.
The purpose in this paper was to prove the existence of four-dimensional, spherically symmetric,
soliton and black hole solutions for any integer value of $N$.
The general approach was briefly outlined in \cite{Baxter1}, but in this paper we have presented
all the detailed results.

We began, in Section \ref{sec:ansatz},
by describing the field equations for ${\mathfrak {su}}(N)$ EYM with a negative cosmological constant,
the ansatz for the metric and gauge potential
and the boundary conditions at the origin, black hole event horizon (if there is one) and infinity.
We considered only purely magnetic gauge fields, which are described by $N-1$ gauge field functions $\omega _{j}$.
We also described how any ${\mathfrak {su}}(2)$ EYM soliton or black hole solution can be embedded
to give an ${\mathfrak {su}}(N)$ EYM solution for any value of $N$.
The goal of the paper is to prove the existence of genuinely ${\mathfrak {su}}(N)$ solutions, that is,
solutions of the ${\mathfrak {su}}(N)$ EYM field equations which are not embedded ${\mathfrak {su}}(2)$ solutions.

The field equations for ${\mathfrak {su}}(N)$ EYM are singular at the origin, black hole event horizon
and at infinity.
In Section \ref{sec:local} we proved local existence of solutions in neighbourhoods of these singular
points, using an approach in \cite{Breitenlohner2,Oliynyk1}.
Our key existence results are in Section \ref{sec:existence}.
Firstly, we are able to prove the existence
of genuinely ${\mathfrak {su}}(N)$ solutions in a neighbourhood of an embedded ${\mathfrak {su}}(2)$
solution, and, as a corollary, the existence of ${\mathfrak {su}}(N)$ solutions for which all the gauge
field functions have no zeros, for any negative value of the cosmological constant.
Secondly, we show that, for any fixed values of the initial parameters at either the origin or the black hole
event horizon, which fix the local solution in a neighbourhood of the relevant starting point,
then for $\left| \Lambda \right| $ sufficiently large, this local solution can be extended to infinity
to give a soliton or black hole solution for which all the gauge field functions have no zeros.
These solutions when $\left| \Lambda \right| $ is large are of particular interest, because at least some of them
are linearly stable under spherically symmetric perturbations.
An outline of this result can be found in \cite{Baxter1}, and the details will be presented elsewhere \cite{Baxter3}.

Our main result is therefore the existence of four-dimensional, spherically symmetric, asymptotically adS,
soliton and black hole solutions
of the ${\mathfrak {su}}(N)$ EYM field equations with $N-1$ gauge field degrees of freedom.
For black holes, this means that there is no limit on the number of gauge field degrees of freedom
(or `hair') with which the black hole may be endowed.
Of course, it is the stability of at least some of these black holes which is central to their importance for the
no-hair conjecture, an issue to which we shall return \cite{Baxter3}.
This result raises the interesting question of whether there are solitons or black holes with an infinite number of gauge
field degrees of freedom.
There is some evidence \cite{Winstanley4} for black hole solutions of ${\mathfrak {su}}(\infty )$ EYM,
but more work is required in this area.

There are a number of generalizations of the model considered here which would merit further study.
We have not considered topological black holes in adS, although solutions are known for the ${\mathfrak {su}}(2)$
gauge group \cite{Radu1}.
For ${\mathfrak {su}}(2)$ EYM in adS, static, axially symmetric soliton \cite{Radu2} and black hole
\cite{Radu3} solutions are known: it is likely that generalizations of these to ${\mathfrak {su}}(N)$ EYM
exist.
Rotating ${\mathfrak {su}}(2)$ EYM black holes in adS have also been found \cite{Radu4}, and again
one would expect generalizations to the larger gauge group to exist.
However, the numerical challenges in finding such solutions cannot be underestimated.
The stability of these solutions also remains an open question.
In this paper we have restricted our attention to purely magnetic gauge fields.
For ${\mathfrak {su}}(2)$ EYM in adS, dyonic regular and black hole solutions exist \cite{Bjoraker},
even though non-trivial ${\mathfrak {su}}(2)$ EYM solutions in asymptotically flat space must be purely magnetic
\cite{Ershov}.
Since ${\mathfrak {su}}(N)$ solutions in asymptotically flat space may have electric as well as magnetic charge
\cite{Galtsov1}, it would not be surprising to find dyonic ${\mathfrak {su}}(N)$ EYM solutions in adS,
although including the electric part of the gauge field will only introduce a single additional gauge field
degree of freedom.
The stability of the dyonic solutions, even in the ${\mathfrak {su}}(2)$ case, remains an open question.
Next, EYM solitons and black holes in more than four space-time dimensions have recently received
much attention in the literature (see \cite{Volkov2} for a review and references), and analogue higher dimensional
solutions of ${\mathfrak {su}}(N)$ EYM would be expected.
Finally, one would expect to be able to extend our results to arbitrary compact gauge group:
in \cite{Oliynyk1} local existence theorems are proved for compact gauge group in asymptotically flat space, and
extending these to include a negative cosmological constant should be possible, leading to existence theorems
along the lines of those proved in Section \ref{sec:existence}.

Having shown that there is no limit to the amount of hair a black hole in adS can be given, the natural question
arises of the consequences of these black hole solutions for gravitational physics.
Firstly, there is the impact on the status of the ``no-hair'' conjecture: these black holes require an unlimited
number of parameters to fully describe their configurations, and therefore are contrary to the ``spirit'' of the
``no-hair'' conjecture, namely that black holes are fundamentally simple objects.
The fact that these black holes with unlimited amounts of hair exist in anti-de Sitter space may be significant,
particularly in view of the adS/CFT (conformal field theory) correspondence \cite{adSCFT}.
It has been conjectured \cite{Hertog1} that there are observables in the dual (deformed) CFT which are
sensitive to the presence of black hole hair,
and an adS/CFT interpretation of some stable seven-dimensional black holes with ${\mathfrak {so}}(5)$
gauge fields is given in \cite{Gauntlett1} (see also \cite{Gubser} for a discussion of non-abelian solutions
in the context of the adS/CFT correspondence).
Black holes with non-abelian gauge fields in supergravity have begun to attract attention recently
\cite{SUGRA} and understanding the supersymmetric analogues of the solutions we have found in this
paper will be an important question to which we plan to return in the near future.

\ack
We thank Brien Nolan for helpful discussions, and Marc Helbling for
writing the code used to produce some of the figures.
The work of JEB is supported by a studentship from EPSRC (UK),
and the work of EW is supported by STFC (UK), grant number PPA/D000351/1.

\end{document}